# Quantum mechanical modeling of anharmonic phonon-phonon scattering in nanostructures

Yangyu Guo [1*], Marc Bescond [2], Zhongwei Zhang [1,3], Mathieu Luisier [4], Masahiro Nomura [1†], Sebastian Volz [1,2‡]

[1] *Institute of Industrial Science, The University of Tokyo, Tokyo 153-8505, Japan*
[2] *LIMMS, CNRS-IIS UMI 2820, The University of Tokyo, Tokyo 153-8505, Japan*
[3] *Center for Phononics and Thermal Energy Science, School of Physics Science and Engineering, Tongji University, 200092 Shanghai, PR China*
[4] *Integrated Systems Laboratory, ETH Zürich, 8092 Zürich, Switzerland*

(Dated: July 28, 2020)

## Abstract

The coherent quantum effect becomes increasingly important in the heat dissipation bottleneck of semiconductor nanoelectronics with the characteristic size shrinking down to few nanometers scale nowadays. However, the quantum mechanical model remains elusive for anharmonic phonon-phonon scattering in extremely small nanostructures with broken translational symmetry. It is a long-term challenging task to correctly simulate quantum heat transport including anharmonic scattering at a scale relevant to practical applications. In this article, we present a clarified theoretical formulation of anharmonic phonon nonequilibrium Green's function (NEGF) formalism for both 1D and 3D nanostructures, through a diagrammatic perturbation expansion and an introduction of Fourier's representation to both harmonic and anharmonic terms. A parallelized computational framework with first-principle force constants input is developed for large-scale quantum heat transport simulation. Some crucial approximations in numerical implementation are investigated to ensure the balance between numerical accuracy and efficiency. A quantitative validation is demonstrated for the anharmonic phonon NEGF formalism and computational framework by modeling cross-plane heat transport through silicon thin film. The phonon-phonon scattering is shown to be appreciable and to introduce about 20% reduction of thermal conductivity at room temperature even for a film thickness around 10 nm. The present methodology provides a robust platform for the device quantum thermal modeling, as well as the study on the transition from coherent to incoherent heat transport in nano-phononic crystals. This work thus paves the way to understand and to manipulate heat conduction via the wave nature of phonons.

---

[*] yyguo@iis.u-tokyo.ac.jp
[†] nomura@iis.u-tokyo.ac.jp
[‡] volz@iis.u-tokyo.ac.jp

## 1. Introduction

The manipulation and control of heat transport in dielectric crystal mediated by phonons has been a significant issue in modern technology and industrial applications. The aim of low phonon thermal conductivity is pursued in thermoelectric [1] and thermal barrier coating materials [2], whereas high phonon thermal conductivity is desired for heat dissipation problems in nanoelectronics [3]. In the past several decades, fruitful efforts have been made in controlling the heat transport by tailoring the mean free path (MFP) of phonons through nanostructures [4-7]. These efforts are mainly based on the particle nature of phonons [8, 9], since the characteristic size of the system is usually much larger than the dominant width (or called coherent length) of phonon wave-packets. With the development of nano-fabrication and manufacturing technology in recent years, heat conduction tuning via the wave nature of phonons based on nano-phononic crystals (PnC) becomes also possible [10-13]. Distinctive behaviors of phonon heat transport have been experimentally demonstrated in this wave regime: the period-controlled lattice thermal conductivity minimum of superlattices (SLs) [14], the order of nano-PnC slowing down the heat propagation [15], nanodot induced Anderson localization of thermal phonons in SLs [16], the aperiodicity of SLs reducing thermal conductivity via multiple localization [17], and so on [18-20].

The modeling of phonon heat transport in this wave-like coherent regime to particle-like incoherent regime remains, however, a challenging task. The widely adopted semi-classical phonon Boltzmann equation becomes no longer valid [21, 22] as it describes only the evolution of phonon-particle population whereas ignoring the crucial phase information (interference effect) of phonon modes in this situation. There has been some theoretical effort to describe such coherent heat transport by a lattice dynamical model [15, 23, 24], with the anharmonic phonon scattering included by an imaginary wave vector component dependent on MFP in a phenomenological way. The molecular dynamics (MD) simulation is another popular approach to study the transition from coherent to incoherent heat transport in nano-PnC such as SLs [25-28]. As MD is a classical method, some essential quantum behavior of phonons would be lost at relatively low temperatures or in extremely small nanostructures. On the other hand, the empirical atomic interaction potential employed in MD might not be sufficiently accurate or even not available sometimes. Therefore, a full quantum mechanical model without any empirical input parameters is highly desired, which is the main focus of this work.

Non-equilibrium Green's function (NEGF) method [29-32] represents such a full quantum mechanical approach which could principally account for the coherent interaction and the incoherent scattering in the same footprint. In comparison to the Boltzmann transport theory, NEGF method describes the evolution of both population and coherences of phonon modes [33] such that it will be able to capture the wave-like to particle-like behaviors of phonon heat transport. The NEGF formalism was originally proposed around the last mid-century [34-36] during the development of quantum field theory [37] mainly for electrons, and has been relatively well established and widely applied in quantum transport modeling of nano-electronic devices [38-41]. Later the NEGF method was introduced to model phonon transport in the early years of this century attributed to several pioneering works [42-45]. Due to the large computational cost, the phonon NEGF has been often applied for ballistic heat transport through relatively simple structures like low-dimensional nanostructures (ultra-thin nanowire [42, 46, 47], nanotubes [45, 48], molecular junctions [49], 2D structures [50-52], *etc*.), interfaces [53-58] and SLs [17, 59-61]. Very few works [43, 44, 62-66] directly take into account the anharmonic phonon-phonon scattering, yet often considering few-atom systems like atomic chain or junction [43, 44, 62, 63], and single-unit-cell interface [66]. An indirect treatment of incoherent phonon scattering was also proposed by the Büttiker probe approach [67-69] inspired by its counterpart in electron NEGF, yet it relies on fitting the anharmonic scattering rates with empirical expressions. The challenge in NEGF modeling of anharmonic heat transport comes from not only the difficulty in numerical implementation, but also the less established theoretical formalism. As will be declared later in Section 2, the crucial self-energy expressions for phonon-phonon scattering remains still diverse in several prototypical literature [43, 44, 64]. As a result, there still lacks a solid quantitative validation of the anharmonic phonon NEGF formalism to the authors' best knowledge. Another important issue in phonon NEGF modeling is the input of force constant (FC) matrix. The empirical atomic interaction potential is usually adopted to extract the harmonic FC matrix [42, 45-47, 49-53, 55, 56] and anharmonic FC matrix [43, 44, 62-65]. In recent years, attributed to the advance in density-functional theory (DFT) [70] and computational power, the first-principle input has been also introduced into phonon NEGF codes, yet mostly harmonic FC matrix in ballistic heat transport [48, 54, 57-61] except a recent work [66] also considering the anharmonic one.

The aim of the present work includes several aspects. Firstly, we aim to present a clarified formulation of the anharmonic phonon NEGF method through a diagrammatic perturbation expansion of Green's function and a thorough comparison to existing results. As a further step, we will extend the formalism to anharmonic heat transport through 3D nanostructures with transverse periodicity by introducing a Fourier's representation to both the harmonic and anharmonic terms. Secondly, we aim to develop a first-principle-based numerical framework for large-scale quantum heat transport simulation by introducing advanced computation techniques as well as DFT input of harmonic and anharmonic FCs. Finally, a quantitative validation of the anharmonic phonon NEGF formalism and numerical framework will be demonstrated by several benchmark studies. The crucial approximations beyond the treatment in a previous work [64] by one of the co-authors will be elucidated to ensure feasible yet still accurate large-scale simulation. The remaining of this article will be arranged as follows: the theoretical formulations and numerical implementation of the phonon NEGF method will be described in Section 2; the validation results and related discussions will be shown in Section 3; concluding remarks are made in Section 4.

## 2. Mathematical and Numerical Models

In this section, we will present a clarified anharmonic phonon NEGF formalism for each relevant physical system. Section 2.1 will firstly focus on 1D nanostructures without periodicity. Then a Fourier's representation is introduced to the anharmonic formalism in Section 2.2 for 3D nanostructures with transverse periodicity, which is a paradigm for quantum heat transport across thin films, interfaces, SLs and multi-layer nanostructures. In Section 2.3, the numerical implementation of the phonon NEGF formalism will be elucidated in detail, including the computational strategies and techniques for large-scale simulation. In Section 2.4, the input of harmonic and anharmonic FC matrices from first-principle (DFT) calculation will be introduced.

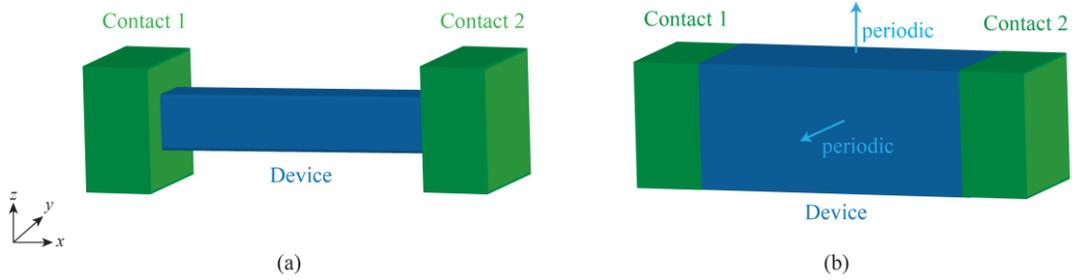

Figure 1. Schematic of the physical model in phonon NEGF formalism: (a) 1D nanostructure; (b) 3D nanostructure with transverse periodicity.

*2.1 Anharmonic phonon NEGF formalism for 1D nanostructures*

In this sub-section, we firstly consider quantum heat transport through a 1D nanostructure without any translational periodicity, as it is shown in Figure 1(a). The total Hamiltonian operator of the system in the Heisenberg representation is expressed as [8]:

$$H(t) = H_0(t) + V(t), \qquad (1)$$

where $H_0(t)$ and $V(t)$ are the exactly solvable harmonic part and the third-order anharmonic perturbation part respectively:

$$H_0(t) = \frac{1}{2}\sum_{n,i} \dot{u}_n^i(t)\dot{u}_n^i(t) + \frac{1}{2}\sum_{nm}\sum_{ij} \Phi_{nm}^{ij} u_n^i(t) u_m^j(t), \qquad (2)$$

$$V(t) = \frac{1}{3!}\sum_{nml}\sum_{ijk} \Phi_{nml}^{ijk} u_n^i(t) u_m^j(t) u_l^k(t), \qquad (3)$$

where $n$, $m$, $l$ denotes atomic index, whereas $i$, $j$, $k$ denotes Cartesian coordinate index ($x$, $y$, $z$). $u_n^i(t)$ is the atomic displacement operator rescaled with atomic mass ($u_n^i = r_n^i\sqrt{M_n}$, with $r_n^i$ the real atomic displacement operator and $M_n$ the atomic mass). The dot on the atomic displacement operator means the time derivative (corresponding to atomic velocity). The normalized second-order and third-order force constant $\Phi_{nm}^{ij}$ and $\Phi_{nml}^{ijk}$ are defined as:

$$\Phi_{nm}^{ij} = \frac{1}{\sqrt{M_n M_m}} \left. \frac{\partial^2 E}{\partial r_n^i \partial r_m^j} \right|_0, \tag{4}$$

$$\Phi_{nml}^{ijk} = \frac{1}{\sqrt{M_n M_m M_l}} \left. \frac{\partial^3 E}{\partial r_n^i \partial r_m^j \partial r_l^k} \right|_0, \tag{5}$$

where $E$ denotes the atomic interaction potential, and the subscript '0' denotes equilibrium position of crystal lattice.

The governing equation for the retarded Green's function of the device in a matrix notation is [43, 64]:

$$\mathbf{G}^R(\omega) = \left[ \omega^2 \mathbf{I} - \mathbf{\Phi} - \mathbf{\Sigma}^R(\omega) \right]^{-1}, \tag{6}$$

where $\mathbf{I}$ denotes the unity matrix, $\mathbf{\Phi}$ denotes the second-order FC matrix with its component given in Eq. (4), and the superscript '−1' means the inverse of a matrix. Note that throughout this work the bold character represents a matrix, whereas its component is denoted by a regular one. $\omega$ represents the frequency of the phonon modes through the system. The retarded self-energy matrix includes the contribution from both the two contacts and the anharmonic phonon-phonon scattering within the device:

$$\mathbf{\Sigma}^R(\omega) = \mathbf{\Sigma}_1^R(\omega) + \mathbf{\Sigma}_2^R(\omega) + \mathbf{\Sigma}_s^R(\omega). \tag{7}$$

The retarded contact self-energy matrix in Eq. (7) is related to the uncoupled retarded Green's function of the contacts ($\mathbf{g}_{1(2)}^R(\omega)$) as [43, 53]:

$$\mathbf{\Sigma}_{1(2)}^R(\omega) = \mathbf{\tau}_{1(2)} \mathbf{g}_{1(2)}^R(\omega) \mathbf{\tau}_{1(2)}^\dagger, \tag{8}$$

with $\mathbf{\tau}_1$ (or $\mathbf{\tau}_2$) denoting the harmonic interaction FC matrix between the device and contacts. For brevity, the subscript '1(2)' is introduced to refer to contact 1 or contact 2 hereafter. The contact self-energy matrix is related to the surface Green's function of the contacts, which are calculated by the decimation technique [32, 71]. The calculation of the retarded scattering self-energy $\mathbf{\Sigma}_s^R(\omega)$ will be introduced later.

For ballistic heat transport problems, usually we only need to resolve the retarded Green's function of the device in Eq. (6) to get the transmission through the system using the Caroli formula [29, 42, 45]. However, for anharmonic heat transport, we also have to compute the greater/lesser Green's function as [43, 64]:

$$\mathbf{G}^{>,<}(\omega) = \mathbf{G}^R(\omega)\mathbf{\Sigma}^{>,<}(\omega)\mathbf{G}^A(\omega),  \tag{9}$$

where the advanced Green's function of the device is the Hermitian conjugate of the retarded Green's function defined as: $\mathbf{G}^A(\omega) = \left(\mathbf{G}^R(\omega)\right)^\dagger$. The greater/lesser self-energy matrix also includes the contribution from both the two contacts and the anharmonic phonon-phonon scattering in the device:

$$\mathbf{\Sigma}^{>,<}(\omega) = \mathbf{\Sigma}_1^{>,<}(\omega) + \mathbf{\Sigma}_2^{>,<}(\omega) + \mathbf{\Sigma}_s^{>,<}(\omega). \tag{10}$$

The greater/lesser contact self-energy matrix in Eq. (10) is related to the uncoupled greater/lesser Green's function of the contacts ($\mathbf{g}_{1(2)}^{>,<}(\omega)$) as:

$$\mathbf{\Sigma}_{1(2)}^{>,<}(\omega) = \mathbf{\tau}_{1(2)} \mathbf{g}_{1(2)}^{>,<}(\omega) \mathbf{\tau}_{1(2)}^\dagger. \tag{11}$$

The uncoupled greater/lesser Green's function of the contacts could be derived from the fluctuation-dissipation theorem at equilibrium [62] and are computed as:

$$\mathbf{g}_{1(2)}^{>}(\omega) = \left[1 + f^{eq}(\omega, T_{1(2)})\right]\left[\mathbf{g}_{1(2)}^R - \left(\mathbf{g}_{1(2)}^R\right)^\dagger\right], \tag{12}$$

$$\mathbf{g}_{1(2)}^{<}(\omega) = f^{eq}(\omega, T_{1(2)})\left[\mathbf{g}_{1(2)}^R - \left(\mathbf{g}_{1(2)}^R\right)^\dagger\right], \tag{13}$$

where $f^{eq}(\omega, T_{1(2)})$ denotes the Bose-Einstein equilibrium phonon distribution at the contact temperature $T_{1(2)}$.

A crucial part of the anharmonic phonon NEGF formalism is the calculation of the greater/lesser scattering self-energy matrix $\mathbf{\Sigma}_s^{>,<}(\omega)$ in Eq. (10). The retarded scattering self-energy matrix in Eq. (7) is then related to the greater/lesser scattering self-energy matrix as [64]:

$$\mathbf{\Sigma}_s^R(\omega) = \frac{1}{2}\left[\mathbf{\Sigma}_s^{>}(\omega) - \mathbf{\Sigma}_s^{<}(\omega)\right] + i\mathrm{P}\int_{-\infty}^{\infty} \frac{d\omega'}{2\pi} \frac{\mathbf{\Sigma}_s^{>}(\omega') - \mathbf{\Sigma}_s^{<}(\omega')}{\omega - \omega'}. \tag{14}$$

The second term on the right-hand side of Eq. (14) is the Cauchy principal integral, and is often neglected for computational simplicity [64]. This term represents the frequency shift of phonon modes due to the anharmonic phonon-phonon scattering and is usually very small for most applications. In terms of the greater/lesser scattering self-energy, we find similar expressions as available in the literature yet with different coefficients [43, 44, 64], as it is summarized in Table 1. For a clear comparison, we keep the notation of third-order FC in the original literature whereas rewrite the notation of other variables (atomic displacement, Green's function and self-energy) to be consistent with that in this work. This discrepancy might be due to the very complicated derivation process of the anharmonic scattering self-energy through the diagrammatic perturbation expansion of phonon Green's function, the details of which are often not provided [43, 44, 64]. Since there is no concretely validated and widely accepted expression of anharmonic scattering self-energy for heat transport in nanostructures, our strategy to resolve this issue includes a twofold procedure: (1) firstly we conduct a diagrammatic perturbation derivation of the self-energy expression by ourselves; (2) then we demonstrate a quantitative validation of the theoretical formalism.

Table 1. Summary of expressions for greater/lesser anharmonic phonon scattering self-energy.

| Third-order term in Hamiltonian | Scattering self-energy | References |
|---|---|---|
| $\sum_{ijk} V_{ijk}^{(3)} u_i u_j u_k$ | $i\Sigma_{s,in}^{<}(\omega) = \hbar \sum_{jklm} \int_{-\infty}^{\infty} V_{ijk}^{(3)} G_{jl}^{<}(\omega') G_{km}^{<}(\omega-\omega') V_{lmn}^{(3)} d\omega'$ | Eq. (27) in Ref. [43] |
| $\frac{1}{3} \sum_{ijk} T_{ijk} u_i u_j u_k$ | $\Sigma_{s,jk}^{<}(\omega) = 2i\hbar \sum_{lmrs} T_{jlm} T_{rsk} \int_{-\infty}^{\infty} G_{lr}^{<}(\omega') G_{ms}^{<}(\omega-\omega') \frac{d\omega'}{2\pi}$ | First term of Eq. (13) in Ref. [44] |
| $\frac{1}{3!} \sum_{ijk} \sum_{lmn} dV_{lmn}^{(3)ijk} u_l^i u_m^j u_n^k$ | $\Sigma_{s,nm}^{>,<ij}(\omega) = 2i\hbar \sum_{k_1 k_2 k_3 k_4} \sum_{l_1 l_2 l_3 l_4} \int_{-\infty}^{\infty} \frac{d\omega'}{2\pi} dV_{nl_1 l_2}^{(3)ik_1 k_2} dV_{l_3 l_4 m}^{(3)k_3 k_4 j}$ $\times G_{l_1 l_3}^{>,<k_1 k_3}(\omega+\omega') G_{l_4 l_2}^{<,>k_4 k_2}(\omega')$ | Eq. (7) in Ref. [64] |
| $\frac{1}{3!} \sum_{nml} \sum_{ijk} \Phi_{nml}^{ijk} u_n^i u_m^j u_l^k$ | $\Sigma_{s,ll'}^{>,<ij}(\omega) = \frac{1}{2} i\hbar \sum_{l_1 l_2 l_3 l_4} \sum_{j_1 j_2 j_3 j_4} \int_{-\infty}^{\infty} \frac{d\omega'}{2\pi} \Phi_{ll_1 l_2}^{ij_1 j_2} \Phi_{l'l_3 l_4}^{jj_3 j_4}$ $\times G_{l_1 l_4}^{>,<j_1 j_4}(\omega') G_{l_2 l_3}^{>,<j_2 j_3}(\omega-\omega')$ | Present |

Through the diagrammatic perturbation expansion, we obtain the expression of the greater/lesser self-energy matrix for third-order anharmonic phonon scattering as:

$$\Sigma_{s,ll'}^{>,<ij}(\omega) = \frac{1}{2} i\hbar \sum_{l_1 l_2 l_3 l_4} \sum_{j_1 j_2 j_3 j_4} \int_{-\infty}^{\infty} \frac{d\omega'}{2\pi} \Phi_{ll_1 l_2}^{ij_1 j_2} \Phi_{l'l_3 l_4}^{jj_3 j_4} G_{l_1 l_4}^{>,<j_1 j_4}(\omega') G_{l_2 l_3}^{>,<j_2 j_3}(\omega-\omega'), \qquad (15)$$

with the detailed derivation process given in Appendix A. In comparison to the expressions of anharmonic scattering self-energy in Ref. [43] and Ref. [64], there exists a factor difference of $\pi/9$ and 4 respectively with respect to the present result of Eq. (15). Considering that the third-order FC ($T_{ijk}$ in Table 1) is the half of the present one ($\Phi_{nml}^{ijk}$), the part of self-energy corresponding to the three-phonon anharmonic scattering in Ref. [44] is actually consistent with the present expression. A further theoretical corroboration of Eq. (15) will be shown in Section 2.2 when we compare it to the expression of anharmonic phonon scattering self-energy in bulk materials. Up to now, Eq. (6), Eq. (9), Eq. (14) and Eq. (15) constitute a closed set of coupled matrix equations, which have to be solved through an iterative process called self-consistent Born approximation (SCBA) [41, 64].

*2.2 Anharmonic phonon NEGF formalism for 3D nanostructures*

In this sub-section, we consider quantum heat transport through 3D nanostructures with transverse periodicity as it is shown in Figure 1(b). Attributed to this periodicity in the transverse direction, the phonon NEGF formalism in Section 2.1 can be rewritten into a Fourier's representation by introducing a transverse wave vector $\mathbf{q}_\perp$ and the following Fourier transform for the Green's function:

$$G_{ll'}^{ij}(\omega) = \frac{1}{N} \sum_{\mathbf{q}_\perp} \exp(i\mathbf{q}_\perp \cdot \Delta\mathbf{R}_\perp) G_{l_x l'_x}^{ij}(\omega; \mathbf{q}_\perp), \qquad (16)$$

$$G_{l_x l'_x}^{ij}(\omega; \mathbf{q}_\perp) = \sum_{\Delta\mathbf{R}_\perp} \exp(-i\mathbf{q}_\perp \cdot \Delta\mathbf{R}_\perp) G_{ll'}^{ij}(\omega), \qquad (17)$$

where $N$ is the number of transverse wave vectors, and $\Delta\mathbf{R}_\perp = (l_y - l'_y)\mathbf{a}_2 + (l_z - l'_z)\mathbf{a}_3$ with $\mathbf{a}_2$ and $\mathbf{a}_3$ the lattice vectors along $y$- and $z$-direction respectively. Here $l \equiv (l_x, l_y, l_z)$ denotes the index of lattice unit cell in the device. Note that in the Green's function (and related variables such as self-energy matrix discussed later), the subscript $l$ also includes the index of atoms within the corresponding lattice unit cell, *i.e.* $G_{ll'}^{ij}(\omega) \equiv G_{l\kappa, l'\kappa'}^{ij}(\omega)$ with $\kappa$ the atomic index in a lattice unit cell. For the sake of clarity, we only keep the index of lattice unit cell.

With the help of Eq. (16) and Eq. (17), the formalism in Section 2.1 can be rewritten into the Fourier's representation and the main governing equations Eq. (6), Eq. (9), Eq. (14) and Eq. (15) become:

$$\mathbf{G}^{\mathrm{R}}(\omega;\mathbf{q}_\perp) = \left[\omega^2 \mathbf{I} - \tilde{\mathbf{\Phi}}(\mathbf{q}_\perp) - \mathbf{\Sigma}^{\mathrm{R}}(\omega;\mathbf{q}_\perp)\right]^{-1}, \tag{18}$$

$$\mathbf{G}^{>,<}(\omega;\mathbf{q}_\perp) = \mathbf{G}^{\mathrm{R}}(\omega;\mathbf{q}_\perp)\mathbf{\Sigma}^{>,<}(\omega;\mathbf{q}_\perp)\mathbf{G}^{\mathrm{A}}(\omega;\mathbf{q}_\perp), \tag{19}$$

$$\mathbf{\Sigma}_{\mathrm{s}}^{\mathrm{R}}(\omega;\mathbf{q}_\perp) = \frac{1}{2}\left[\mathbf{\Sigma}_{\mathrm{s}}^{>}(\omega;\mathbf{q}_\perp) - \mathbf{\Sigma}_{\mathrm{s}}^{<}(\omega;\mathbf{q}_\perp)\right] + i\mathrm{P}\int_{-\infty}^{\infty}\frac{d\omega'}{2\pi}\frac{\mathbf{\Sigma}_{\mathrm{s}}^{>}(\omega';\mathbf{q}_\perp) - \mathbf{\Sigma}_{\mathrm{s}}^{<}(\omega';\mathbf{q}_\perp)}{\omega - \omega'}, \tag{20}$$

$$\Sigma_{\mathrm{s},l_xl_x'}^{<,>ij}(\omega;\mathbf{q}_\perp) = \frac{1}{2}i\hbar \sum_{l_{1x}l_{2x}l_{3x}l_{4x}} \sum_{j_1j_2j_3j_4} \frac{1}{N}\sum_{\mathbf{q}_\perp'} \tilde{\Phi}_{l_xl_{1x}l_{2x}}^{ij_1j_2}(\mathbf{q}_\perp',\mathbf{q}_\perp-\mathbf{q}_\perp')\tilde{\Phi}_{l_x'l_{3x}l_{4x}}^{jj_3j_4}(\mathbf{q}_\perp'-\mathbf{q}_\perp,-\mathbf{q}_\perp')$$
$$\times \int_{-\infty}^{\infty}\frac{d\omega'}{2\pi}G_{l_{1x}l_{4x}}^{<,>j_1j_4}(\omega';\mathbf{q}_\perp')G_{l_{2x}l_{3x}}^{<,>j_2j_3}(\omega-\omega';\mathbf{q}_\perp-\mathbf{q}_\perp') \tag{21}$$

In Eq. (18), the Fourier's representation of the harmonic FC matrix is defined as:

$$\tilde{\Phi}_{l_xl_x'}^{ij}(\mathbf{q}_\perp) = \sum_{\Delta\mathbf{R}_\perp}\Phi_{ll'}^{ij}\exp(-i\mathbf{q}_\perp\cdot\Delta\mathbf{R}_\perp), \tag{22}$$

which is consistent with the definition in the previous ballistic NEGF formalism [53, 58]. In Eq. (21), the Fourier's representation of the third-order anharmonic FC matrix is defined as:

$$\tilde{\Phi}_{l_xl_{1x}l_{2x}}^{ij_1j_2}(\mathbf{q}_\perp,\mathbf{q}_\perp') = \sum_{\Delta\mathbf{R}_\perp}\sum_{\Delta\mathbf{R}_\perp'}\Phi_{ll_1l_2}^{ij_1j_2}\exp(-i\mathbf{q}_\perp\cdot\Delta\mathbf{R}_\perp)\exp(-i\mathbf{q}_\perp'\cdot\Delta\mathbf{R}_\perp'), \tag{23}$$

with $\Delta\mathbf{R}_\perp = (l_y - l_{1y})\mathbf{a}_2 + (l_z - l_{1z})\mathbf{a}_3$, $\Delta\mathbf{R}_\perp' = (l_y - l_{2y})\mathbf{a}_2 + (l_z - l_{2z})\mathbf{a}_3$ here. The third-order FC matrix has a dependence on two transverse wave vectors since the three-body interaction depends on two relative displacements ($\Delta\mathbf{R}_\perp$ and $\Delta\mathbf{R}_\perp'$) in the cross-sectional direction. The detailed derivation of Eq. (21) with the aid of Eq. (23) is provided in Appendix B. In a recent contribution to the anharmonic phonon NEGF formalism for 3D interfaces [66], a different third-order tensor Fourier's decomposition using the *P* matrix is developed for the anharmonic FC matrix. The authors obtain the anharmonic phonon scattering self-energy matrix below [66]:

$$\Sigma_{\mathrm{s},ur}^{<,>}(\omega;\mathbf{q}_\perp) = i\hbar\sum_{vwpq}\sum_{\mathbf{q}_\perp'}\tilde{V}_{uvw}(\mathbf{q}_\perp,\mathbf{q}_\perp')\tilde{W}_{pqr}(\mathbf{q}_\perp,\mathbf{q}_\perp')\int_{-\infty}^{\infty}d\omega'G_{qv}^{<,>}(\omega';\mathbf{q}_\perp')G_{wp}^{<,>}(\omega-\omega';\mathbf{q}_\perp'), \tag{24}$$

where we have kept most of the notations therein. In spite of a similar mathematical form, our result of Eq. (21) is different from Eq. (24) in terms of the following three aspects: (1) a factor difference since their development is based on the anharmonic phonon scattering self-energy in Ref. [43], as is already shown in Table 1; (2) both the energy conservation ($\omega = \omega' + (\omega - \omega')$) and the momentum conservation ($\mathbf{q}_\perp = \mathbf{q}_\perp' + (\mathbf{q}_\perp - \mathbf{q}_\perp')$) are automatically satisfied in our expression (as it is explicitly shown in the Feynman diagram in Figure 2) whereas only the energy conservation is ensured in Eq. (24); (3) a normalization over the number of transverse wave vectors is also included before the sum over $\mathbf{q}'_\perp$ in our formulation.

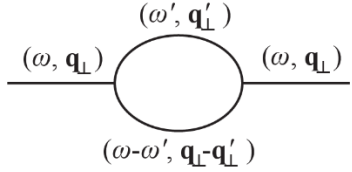

Figure 2. Feynman diagram for three-phonon anharmonic scattering in 3D nanostructures with transverse periodicity.

The expression of anharmonic phonon-phonon scattering self-energy in Eq. (21) can be further corroborated by its counterpart in bulk material. In comparison to the present 3D nanostructures with transverse periodicity, the bulk material also has periodicity in the transport direction. Through a similar diagrammatic perturbation expansion of the phonon Green's function as that in Appendix A, one could obtain the self-energy for three-phonon anharmonic scattering in bulk material as follows [72]:

$$\Sigma_s^<(\omega;\mathbf{q}) = 2i\hbar \sum_{\mathbf{q}_1\mathbf{q}_2} \int_{-\infty}^{\infty} \frac{d\omega_1}{2\pi} \int_{-\infty}^{\infty} \frac{d\omega_2}{2\pi} 2\pi\delta(\omega - \omega_1 - \omega_2)|F(-\mathbf{q},\mathbf{q}_1,\mathbf{q}_2)|^2 G^{(0)<}(\omega_1;\mathbf{q}_1) G^{(0)<}(\omega_2;\mathbf{q}_2), \qquad (25)$$

where $\mathbf{q}$ is the 3D wave vector in bulk material, $|F(-\mathbf{q},\mathbf{q}_1,\mathbf{q}_2)|^2 = F(-\mathbf{q},\mathbf{q}_1,\mathbf{q}_2) \times F^*(-\mathbf{q},\mathbf{q}_1,\mathbf{q}_2)$ with $F^*(-\mathbf{q},\mathbf{q}_1,\mathbf{q}_2) = F(\mathbf{q},-\mathbf{q}_1,-\mathbf{q}_2)$. Here $F(\mathbf{q}_1,\mathbf{q}_2,\mathbf{q}_3)$ is the third-order anharmonic dynamic matrix contributing to the anharmonic term in the Hamiltonian as [72]:

$$V = \frac{1}{3} \sum_{\mathbf{q}_1\mathbf{q}_2\mathbf{q}_3} F(\mathbf{q}_1,\mathbf{q}_2,\mathbf{q}_3) A_{\mathbf{q}_1} A_{\mathbf{q}_2} A_{\mathbf{q}_3}, \qquad (26)$$

where $A_\mathbf{q} = \sqrt{\hbar/2\omega_\mathbf{q}}(a_\mathbf{q} + a^\dagger_{-\mathbf{q}})$ is the phonon normal coordinate operator, with $a_\mathbf{q}, a^\dagger_\mathbf{q}$ the phonon destruction and creation operator respectively. Due to the translational symmetry of the 3D lattice, $F(\mathbf{q}_1,\mathbf{q}_2,\mathbf{q}_3)$ includes a delta function as $\Delta(\mathbf{q}_1+\mathbf{q}_2+\mathbf{q}_3)$ which will be non-vanishing only when $\mathbf{q}_1+\mathbf{q}_2+\mathbf{q}_3 = 0$ or a reciprocal lattice vector [72, 73]. Thus Eq. (25) can be rewritten into:

$$\Sigma_s^<(\omega;\mathbf{q}) = 2i\hbar \sum_{\mathbf{q}_1} \int_{-\infty}^{\infty} \frac{d\omega_1}{2\pi} F(-\mathbf{q},\mathbf{q}_1,\mathbf{q}-\mathbf{q}_1) F(\mathbf{q},-\mathbf{q}_1,\mathbf{q}_1-\mathbf{q}) \mathbf{G}^{(0)<}(\omega_1;\mathbf{q}_1) \mathbf{G}^{(0)<}(\omega-\omega_1;\mathbf{q}-\mathbf{q}_1). \tag{27}$$

Considering the different notations in the third-order term of the Hamiltonian between Eq. (26) and Eq. (3) (a factor 1/3 versus 1/3!), Eq. (27) will be exactly consistent with Eq. (21) once the periodicity of transport direction is released. Note that the normalization factor $1/N$ is included in $|F(-\mathbf{q},\mathbf{q}_1,\mathbf{q}_2)|^2$ [72].

Finally, some general symmetry relations between the non-equilibrium phonon Green's functions for 3D nanostructures with transvese peridocity are demonstrated:

$$\left[\mathbf{G}^>(\omega;\mathbf{q}_\perp)\right]^T = \mathbf{G}^<(-\omega;-\mathbf{q}_\perp), \tag{28}$$

$$\left[\mathbf{G}^<(\omega;\mathbf{q}_\perp)\right]^\dagger = -\mathbf{G}^<(\omega;\mathbf{q}_\perp), \tag{29}$$

$$\left[\mathbf{G}^R(\omega;\mathbf{q}_\perp)\right]^* = \mathbf{G}^R(-\omega;-\mathbf{q}_\perp), \tag{30}$$

$$\left[\mathbf{G}^R(\omega;\mathbf{q}_\perp)\right]^\dagger = \mathbf{G}^A(\omega;\mathbf{q}_\perp). \tag{31}$$

The subscript 'T' in Eq. (28) denotes the transpose of a matrix. We provide a detailed proof of Eq. (28) in Appendix C. The other relations could be proved in a similar way and are not shown here. A comparison to similar general relations between the phonon Green's function for 1D nanostructures [62] is summarized in Table 2.

Table 2. General relations between the non-equilibrium phonon Green's functions for 1D and 3D nanostructures.

| 1D nanostructures [62] | 3D nanostructures (Present result) |
|---|---|
| $\left[\mathbf{G}^>(\omega)\right]^T = \mathbf{G}^<(-\omega)$ | $\left[\mathbf{G}^>(\omega;\mathbf{q}_\perp)\right]^T = \mathbf{G}^<(-\omega;-\mathbf{q}_\perp)$ |
| $\left[\mathbf{G}^<(\omega)\right]^\dagger = -\mathbf{G}^<(\omega)$ | $\left[\mathbf{G}^<(\omega;\mathbf{q}_\perp)\right]^\dagger = -\mathbf{G}^<(\omega;\mathbf{q}_\perp)$ |
| $\left[\mathbf{G}^R(\omega)\right]^* = \mathbf{G}^R(-\omega)$ | $\left[\mathbf{G}^R(\omega;\mathbf{q}_\perp)\right]^* = \mathbf{G}^R(-\omega;-\mathbf{q}_\perp)$ |
| $\left[\mathbf{G}^R(\omega)\right]^\dagger = \mathbf{G}^A(\omega)$ | $\left[\mathbf{G}^R(\omega;\mathbf{q}_\perp)\right]^\dagger = \mathbf{G}^A(\omega;\mathbf{q}_\perp)$ |

*2.3 Numerical implementation*

The numerical implementation of the anharmonic phonon NEGF formalism by the SCBA iterative solution of Eqs. (18)-(21) is challenging for large-scale simulations because of the computational time cost and memory cost. The large time cost is mainly due to the intensive calculation of multiple summations and integrations in the anharmonic scattering self-energy in Eq. (21). Furthermore, the memory cost and time cost related to the matrix storage and operation will be proportional to $(N_xN_d)^2$ and $(N_xN_d)^3$ respectively if the full matrix of the device is directly resolved [39]. Here the device consists of $N_d$ slabs as shown in Figure 3, with the matrix size for each slab being $N_x$. Two advanced computational techniques are introduced in this sub-section to retrieve the situation: (1) the recursive algorithm in Section 2.3.1; (2) the parallelization scheme in Section 2.3.2. We will also briefly introduce the macroscopic variable calculation in Section 2.3.3.

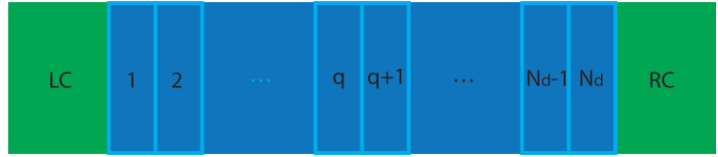

Figure 3. Schematic of recursive algorithm for numerical implementation of phonon NEGF. 'LC' and 'RC' denote left contact (contact 1) and right contact (contact 2) respectively, and the device is divided into $N_d$ slabs.

2.3.1 Recursive algorithm

The recursive algorithm was originally proposed in electron NEGF codes for large-scale nano-electronic device modeling [74], and has been also applied in anharmonic phonon NEGF modeling of heat transport in nanowires [64]. Similar variant of recursive algorithm was also developed in ballistic phonon NEGF [55]. The motivation of the recursive algorithm arises from the fact that only the diagonal blocks and first off-diagonal blocks of Green's functions are required to compute the relevant macroscopic observables (*i.e.* local density and current) of the system. The recursive algorithm only deals with matrices for each slab and its adjacent slab such that the memory cost and time cost will be linearly proportional to the system size along the transport direction as $(N_x)^2N_d$ and $(N_x)^3N_d$ respectively [39]. In this work, we adopt the recursive algorithm [39, 74] to compute the $N_x \times N_x$ diagonal blocks and first off-diagonal blocks of the retarded Green's function and the greater/lesser Green's functions. We also make other approximations in the numerical implementation as follows: (I) only the interaction within neighboring slabs in Figure 3 is considered for harmonic FC; (II) only the atomic nearest-neighbor interaction is considered for third-order anharmonic FC; (III) only the $N_x \times N_x$ diagonal blocks of anharmonic scattering self-energy are considered, and within each diagonal block only the dominant terms between each atom and its nearest-neighbor atoms are computed; (IV) the anharmonic phonon-phonon scattering is considered only in the device region whereas the contacts are harmonic. The approximation (II) has also been adopted in a previous work [64] based on the empirical valence-force-field model, and shall be more or less reasonable for weakly anharmonic materials like silicon considered in this work, as it will be shown later in Section 2.4. In principle, more neighbors should be considered in the third-order FC for strongly anharmonic materials like oxides, yet it remains computationally very challenging for large-scale simulations. Concerning approximation (III), an even simpler approximation is assumed in Ref. [64] where only the 3×3 diagonal blocks of anharmonic scattering self-energy matrix are computed. We will show in Section 3.3 that the approximation in Ref. [64] will significantly overestimate the anharmonic scattering rate and thus underestimate the thermal conductance. In terms of approximation (IV), it has been shown that the harmonic or anharmonic contact has a negligibly small influence on the thermal transport properties [64].

2.3.2 Parallelization scheme

With multiple-CPU (central processing unit) computer facility, the iterative solution of Eqs. (18)-(21) can be parallelized based on the message-passing-interface (MPI) standard [75]. The MPI parallelization scheme has been widely adopted in electron NEGF for large-scale device simulations [41, 76], and is also used in anharmonic phonon NEGF in a previous work [64] by one of the co-authors. In comparison to the parallelization scheme in Ref. [64] for nanowire (1D nanostructures) simulations which require data exchange between different frequency points as shown in Eq. (15), here we also have to exchange the data between different transverse wave vectors for simulation of 3D nanostructures as shown in Eq. (21). Thus, we build a dual-level parallelization scheme for the transverse wave vector and frequency points, both of which are uniformly discretized as it is shown in Figure 4. In the first level, all the transverse wave vectors are parallelly treated, *i.e.* one wave vector will be allocated to each CPU. In the second level, the frequency points are divided into several intervals which are parallelly treated, *i.e.* each CPU will receive a segment with several frequency points. For ballistic heat transport simulation, the problem is embarrassingly parallelized, *i.e.* the retarded Green's function and transmission of each mode $(\omega, \mathbf{q}_\perp)$ is independently calculated in each CPU without the need of data exchange during computation. For anharmonic heat transport simulation, the situation is more complicated as the Green's function of one mode $(\omega, \mathbf{q}_\perp)$ is coupled with many other modes as it is inferred from Eq. (21). Since the phonon Green's functions for each mode are often distributed in different CPUs, data exchange is needed when computing the anharmonic scattering self-energy. We design an algorithm for the data exchange, the details of which are provided in Appendix D.

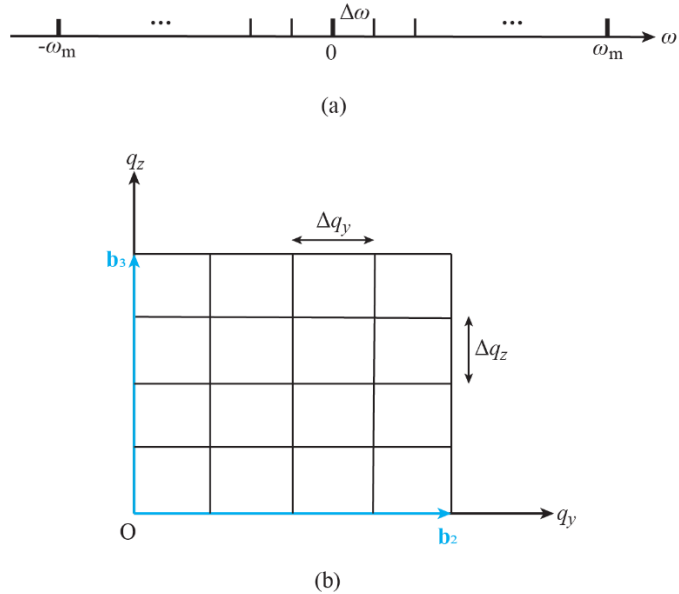

Figure 4. Schematic of the discretization of (a) frequency and (b) transverse wave vector in phonon NEGF simulation. The frequency interval is computed as: $\Delta\omega = 2\omega_m/(N_m-1)$, with $\omega_m$ the maximum crystal phonon frequency and $N_m$ the number of frequency points. The transverse reciprocal lattice vectors are defined as: $\mathbf{b}_i \cdot \mathbf{a}_j = 2\pi\delta_{ij}$, where $i, j = 2, 3$.

### 2.3.3 Macroscopic variable calculation

The local heat flow from the $q^{th}$ slab to $(q+1)^{th}$ slab shown in Figure 3 is related to the first off-diagonal blocks of the lesser phonon Green's function [64] and can be computed as:

$$J_{q \to q+1} = -\sum_{n \in q} \sum_{m \in q+1} \sum_{ij} \frac{1}{N} \sum_{\mathbf{q}_\perp} \int_0^\infty \frac{\hbar\omega}{2\pi} \left[ \tilde{\Phi}_{nm}^{ij}(\mathbf{q}_\perp) G_{mn}^{<,ji}(\omega;\mathbf{q}_\perp) - G_{nm}^{<,ij}(\omega;\mathbf{q}_\perp) \tilde{\Phi}_{mn}^{ji}(\mathbf{q}_\perp) \right]. \tag{32}$$

Eq. (32) can be written as: $J_{q \to q+1} = \frac{1}{N} \sum_{\mathbf{q}_\perp} \int_0^\infty \frac{\hbar\omega}{2\pi} J(\omega;\mathbf{q}_\perp)$, with the heat flow contributed by each mode expressed as:

$$J(\omega;\mathbf{q}_\perp) = -\sum_{n \in q} \sum_{m \in q+1} \sum_{ij} \left[ \tilde{\Phi}_{nm}^{ij}(\mathbf{q}_\perp) G_{mn}^{<,ji}(\omega;\mathbf{q}_\perp) - G_{nm}^{<,ij}(\omega;\mathbf{q}_\perp) \tilde{\Phi}_{mn}^{ji}(\mathbf{q}_\perp) \right]. \tag{33}$$

The local heat flow distribution is used as a criterion of convergence of the SCBA scheme. After each iteration, the local heat flow is computed based on Eq. (32). The SCBA convergence is reached when the conservation of heat flow is fullfilled along the transport direction, *i.e.* the local heat flow is the same for all the slabs within a certain allowed numerical error (1% in the present work).

For purely coherent heat transport, the definition of local temperature is usually not relevant. However, when considering anharmonic phonon-phonon scattering, the local temperature can be computed based on the local energy conservation condition within one slab:

$$\int_0^\infty \rho(\varepsilon, \mathbf{R}_n) \frac{2\hbar\omega}{\hbar^2} \hbar\omega \frac{d(\hbar\omega)}{2\pi} = \int_0^\infty \text{LDOS}(\varepsilon, \mathbf{R}_n) f^{eq}(\omega, T_{eff}) \frac{2\hbar\omega}{\hbar^2} \hbar\omega \frac{d(\hbar\omega)}{2\pi}, \tag{34}$$

where $\mathbf{R}_n$ denotes the spatial position of the $q^{th}$ slab, and $\varepsilon \equiv \omega^2$ is the eigen-value of the harmonic FC matrix. $f^{eq}(\omega, T_{eff})$ is the Bose-Einstein equilibrium phonon distribution at an effective local temperature $T_{eff}$. The local phonon number density in Eq. (34) is related to the diagonal blocks of the lesser phonon Green's function [77] and is computed as:

$$\rho(\varepsilon, \mathbf{R}_n) = \text{Tr}\left[\frac{1}{N}\sum_{\mathbf{q}_\perp} i\mathbf{G}^<_{nn}(\omega; \mathbf{q}_\perp)\right], \tag{35}$$

with 'Tr' denoting the trace of a square matrix, and the subscript '*nn*' represents the sub-block of $\mathbf{G}^<$ corresponding to the q$^{\text{th}}$ slab. The local density of states (LDOS) in Eq. (34) is defined as: $\text{LDOS}(\varepsilon, \mathbf{R}_n) = \text{Tr}[\mathbf{A}_{nn}(\omega)]$, with the spectral function matrix computed as:

$$\mathbf{A}_{nn}(\omega) = \frac{1}{N}\sum_{\mathbf{q}_\perp} \mathbf{A}_{nn}(\omega; \mathbf{q}_\perp) = \frac{1}{N}\sum_{\mathbf{q}_\perp} i\left[\mathbf{G}^>_{nn}(\omega; \mathbf{q}_\perp) - \mathbf{G}^<_{nn}(\omega; \mathbf{q}_\perp)\right]. \tag{36}$$

In a previous work on coupled electron-phonon NEGF modeling [77], a different definition of local temperature was adopted based on the local phonon number conservation condition. This approach amounts to removing one $\hbar\omega$ in the integration on both sides of Eq. (34). Here we keep the convention of using local energy density to characterize the local temperature in non-equilibrium transport as currently done in the heat transport community [21, 67-69].

*2.4 First-principle input*

In this work, the material properties of silicon are adopted for the nanostructures and are obtained by first-principle (DFT) calculation. The DFT calculation is implemented in the open-source package Quantum Espresso (QE) [78] with norm-conserving type pseudo-potential and the LDA exchange-correlation functional for silicon. A kinetic energy cutoff of 60Ry is used for the wave function and a self-consistent convergence threshold of $10^{-12}$ is adopted after independence check. Firstly, a relaxation process on a primitive unit cell is run to obtain an optimized lattice constant of 5.4018Å with an electronic wave vector grid of 8×8×8. For the harmonic FC, the finite displacement method is used as it is implemented in the open-source package Phonopy [79] combined with the DFT calculation in QE. A supercell of 3×3×3 conventional unit cells (totally 216 atoms) is considered, and all the interactions within the supercell are included. An electronic wave vector of 2×2×2 is used in DFT calculation. For the third-order anharmonic FC, the finite displacement method is also used as it is implemented in the open-source package Thirdorder [80] combined with the DFT calculation in QE. A supercell of 2×2×2 conventional unit cells (totally 64 atoms) is adopted. In terms of the atomic interaction range, we consider respectively the first, second and third nearest-neighbor shell to compare the results. An electronic wave vector of 1×1×1 is taken in DFT calculation. After extracting both the harmonic and third-order anharmonic FCs, we use them to calculate the bulk thermal conductivity of silicon (with nature isotope abundancy) in the open-source package ShengBTE [80]. The results at room temperature (300K) calculated (based on the primitive unit cell) with a phonon wave vector 24×24×24 are respectively 120.69, 136.46 and 147.13 (W/m·K) when the atomic interaction within the first, second and third nearest-neighbor shell are separately considered in calculating the third-order FC. The thermal conductivity result is very close to the experimental value (148W/m·K) when the third nearest-neighbor shell is considered, which demonstrates the good quality of DFT harmonic and anharmonic FCs. Due to the challenge of large computational cost, we consider only the first nearest-neighbor shell for the third-order FC, which will underestimate the room-temperature thermal conductivity by about 20%. Since the present work is mainly focused on the demonstration of the anharmonic phonon NEGF methodology, such a simplified treatment captures the dominant third-order anharmonic interaction and is acceptable from the perspective of microscopic modeling.

The supercells have a limited size in the DFT calculation of harmonic and anharmonic FCs. The FC matrices obtained for the DFT supercells are then used to reconstruct the larger FC matrices for nanostructures as input into the large-scale phonon NEGF simulation. In our simulation, one conventional unit cell of silicon is chosen as one slab of the device in Figure 3. The basic idea and procedure of the reconstruction of harmonic FC matrix is shown in Figure 5. Firstly, all the harmonic FC matrices within the central device region, and between the device region and the transverse periodic units $\boldsymbol{\Phi}_0$ ~ $\boldsymbol{\Phi}_8$ are reconstructed from the elementary interaction FC matrices within the supercell. The harmonic FC matrix in Eq. (18) is then computed based on Eq. (22) as:

$$\tilde{\boldsymbol{\Phi}}(\mathbf{q}_\perp) = \boldsymbol{\Phi}_4 + \sum_{i=0,8} \boldsymbol{\Phi}_i \exp\left[\mp i(\mathbf{a}_2 + \mathbf{a}_3)\cdot\mathbf{q}_\perp\right] + \sum_{i=1,7} \boldsymbol{\Phi}_i \exp(\mp i\mathbf{a}_2 \cdot \mathbf{q}_\perp) \\ + \sum_{i=3,5} \boldsymbol{\Phi}_i \exp(\mp i\mathbf{a}_3 \cdot \mathbf{q}_\perp) + \sum_{i=2,6} \boldsymbol{\Phi}_i \exp\left[\mp i(\mathbf{a}_2 - \mathbf{a}_3)\cdot\mathbf{q}_\perp\right],$$

(37)

where the '−' and '+' signs respectively correspond to the first and second index of FC matrix in the summation. The schematic procedure of the reconstruction of third-order FC matrix is shown in Figure 6. The treatment is slightly different because the atomic interaction range for third-order FC is very short. From the elementary interaction matrix in the supercell, we extract the third-order FCs for each atom in a unit cell and its neighboring atoms, *i.e.* a basic third-order FC matrix $\tilde{\Phi}^{ij_1j_2}_{l_x l_{1x} l_{2x}, 0}(\mathbf{q}_\perp, \mathbf{q}'_\perp)$ for one unit cell (one slab) is obtained. When any of the neighboring atoms lies within the transverse periodic unit cell, a phase factor shall be added based on the Fourier's representation of third-order FC matrix in Eq. (23). The third-order FC matrix for the whole device is constructed simply by repeating $\tilde{\Phi}^{ij_1j_2}_{l_x l_{1x} l_{2x}, 0}(\mathbf{q}_\perp, \mathbf{q}'_\perp)$ for times equal to the number of slabs. Note that for heterogenous material properties, the procedure of the reconstruction of device FC matrices shall be slightly adapted to consider the local variation of atomic interaction, for instance in a recent DFT-based ballistic phonon NEGF modeling of disordered lithiated molybdenum disulfide (MoS2) nanostructures [81].

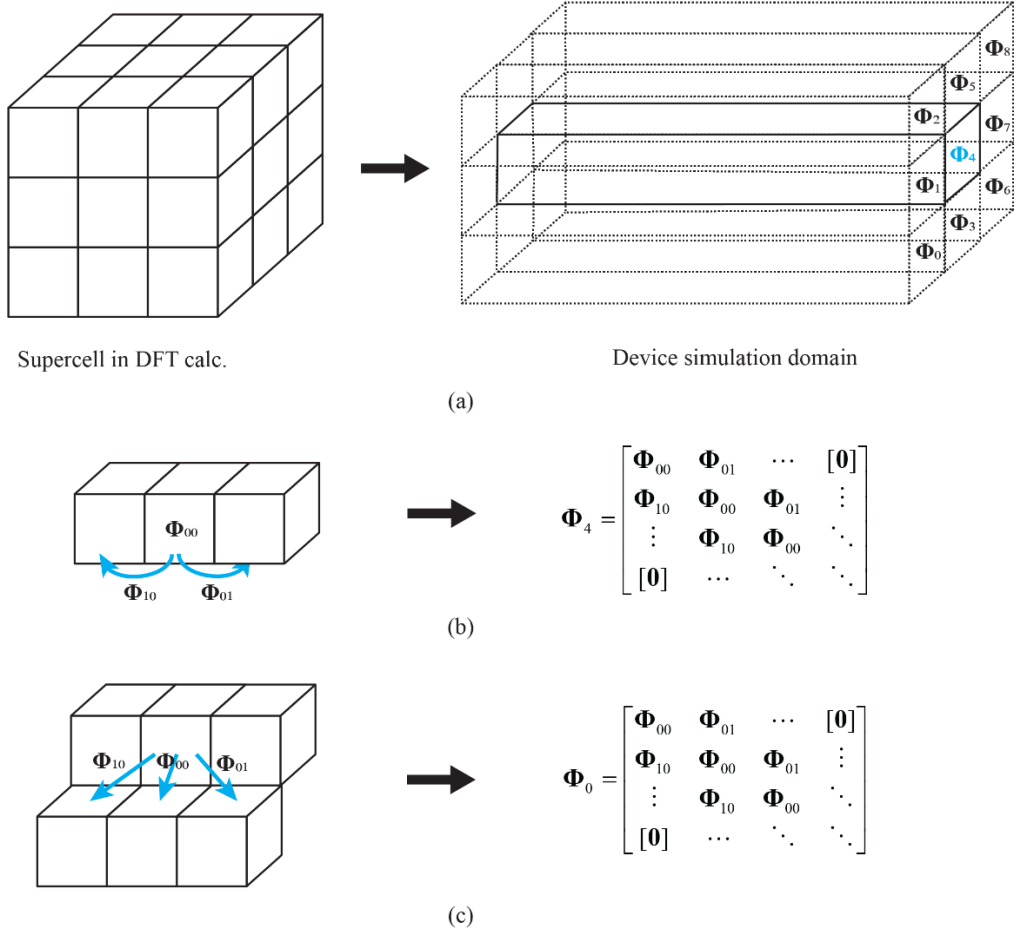

Figure 5. Reconstruction of the harmonic FC matrices as input into the phonon NEGF simulation: (a) schematic of the supercell in DFT calculations, and the device simulation domain, where the solid cuboid represents the central device region whereas the dashed cuboid represents the transverse periodic units that have interaction with the device region with the interaction FC matrix denoted by $\Phi_0 \sim \Phi_8$; (b) reconstruction of the harmonic FC matrix $\Phi_4$ within the central region from the elementary harmonic interaction matrices in the DFT supercell; (c) reconstruction of the representative harmonic FC matrix $\Phi_0$ (between the central region and the transverse periodic unit) from the elementary harmonic interaction matrices in the DFT supercell. Other harmonic FC matrices can be reconstructed similarly. Each cube represents a conventional unit cell.

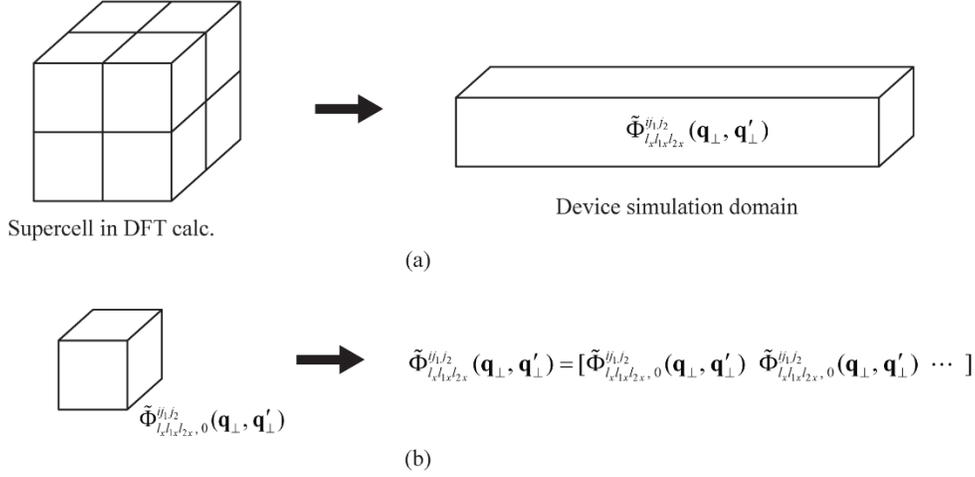

Figure 6. Reconstruction of the third-order anharmonic FC matrix as input into the phonon NEGF simulation: (a) schematic of the supercell in DFT calculation, and the device simulation domain with the corresponding third-order FC matrix $\tilde{\Phi}^{ij_1j_2}_{l_x l_{1x} l_{2x}}(\mathbf{q}_\perp, \mathbf{q}'_\perp)$; (b) reconstruction of the third-order FC matrix from the elementary third-order FC matrix for one unit cell $\tilde{\Phi}^{ij_1j_2}_{l_x l_{1x} l_{2x}, 0}(\mathbf{q}_\perp, \mathbf{q}'_\perp)$ (as explained in the main text) extracted from the DFT supercell. Each cube represents a conventional unit cell, and the parallelpipe denotes the device region.

## 3. Results and Discussions

In this section, we mainly aim to demonstrate the validity of the phonon NEGF formalism and computational framework introduced in Section 2. Firstly, a simple case of ballistic heat transport through a Si/Ge interface is considered in Section 3.1. The quantitative validation of the anharmonic phonon NEGF formalism remains challenging due to its limited computational capability to extremely small structures and/or the usual empirical anharmonic FC input in previous works. In Section 3.2, we tackle this challenge by modeling anharmonic heat transport across a silicon thin film with a thickness larger than 10 nm through our DFT-based large-scale NEGF simulation. Finally, we discuss some crucial approximations in the treatment of anharmonic phonon scattering self-energy in Section 3.3.

*3.1 Validation: ballistic heat transport*

We consider ballistic heat transport through a Si/Ge interface as shown in Figure 7(a), with only one unit cell in the device region. The lattice constant and harmonic FC of Ge are assumed the same as those of Si, with only the atomic mass difference taken into account, to be consistent with already reported studies [54, 57]. This approximate treatment is rather reasonable since the lattice structure of Ge is indeed very similar to that of Si. After numerical convergence test, a frequency mesh of $N_m=201$ and transverse wave vector mesh of $20\times20$ are adopted in the present NEGF simulation with the effect of anharmonic phonon-phonon scattering turned off. Note that the mesh of both frequency and transverse wave vector for the present Si/Ge interface transport shall be much denser than that in homogenous Si film to be discussed in sub-section 3.2. This is due to the very different cut-off frequencies (almost two-fold difference) in the phonon bandstructures of Si and Ge. The ballistic transmission is calculated based on the Caroli formula [29, 54]. The spectral transmission and transmissivity through the Si/Ge interface are shown in Figure 7(b) and (c) respectively, which demonstrates an overall good agreement with respect to previous studies [54, 57]. The minor difference may arise from the slightly different DFT harmonic FCs calculated by different groups.

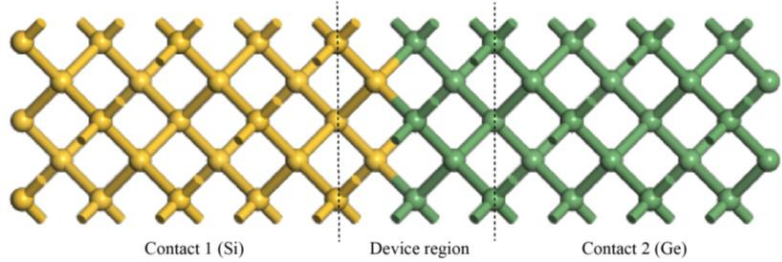

(a)

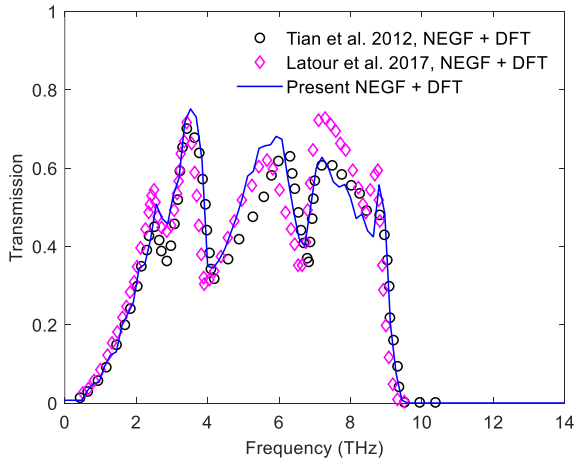

(b)

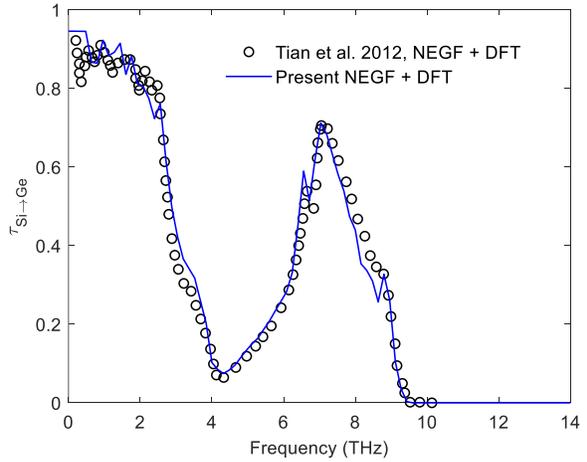

(c)

Figure 7. Phonon NEGF simulation of ballistic heat transport through Si/Ge interface: (a) schematic of numerical model; (b) frequency-dependent interfacial transmission; (c) frequency-dependent interfacial transmissivity (interfacial transmission divided by the transmission of pure Si). The circles and diamonds represent the results of ballistic phonon NEGF with DFT input from Ref. [54] and Ref. [57] respectively, whereas the solid line represents the result of present work.

*3.2 Validation: anharmonic heat transport*

In this sub-section, we present a validation of the theoretical model of anharmonic scattering self-energy in the phonon NEGF formalism. A classical case of heat transport across silicon thin film around 300K is considered, as shown in Figure 8. The temperatures of left contact (contact 1) and right contact (contact 2) are set at 305K and 295K respectively (with a temperature difference $\Delta T = 10K$). The heat transport under such a temperature difference lies still within a linear regime based on our numerical test under a tiny temperature difference of 0.1K. A frequency mesh of 101 and transverse wave vector mesh 6×6 is adopted after numerical convergence test as it is summarized in Table 3. Denser meshes produce thermal conductance within about 2% variation comparing to the present mesh. We consider a series of thickness for the silicon thin film up to 24uc (13 nm) due to computational limitations. For all the cases, the SCBA iterative solution of the governing equations in Section 2.2 converges within 10 iterations. Usually more iterations is needed for larger thickness due to stronger anharmonic phonon-phonon scattering. The computational time cost is about 1h ~ 36h for all the cases in this subsection with 3636 CPUs. The position-dependent spectral heat flow and heat flow across the thin film with a thickness of 20uc (10.8nm) are shown in Figure 9, which demonstrates the good validity of heat flow conservation along the transport direction. After the SCBA convergence, the thermal conductance of the thin film is computed by: $G = J/(A_c \cdot \Delta T)$, with $J$ the average heat flow across the thin film and $A_c$ the cross-section area of the device region ($A_c = \mathbf{a}_2 \cdot \mathbf{a}_3$ here). The effective thermal conductivity of thin film is then related to the thermal conductance as: $\kappa_{\text{eff}} = G \cdot d$.

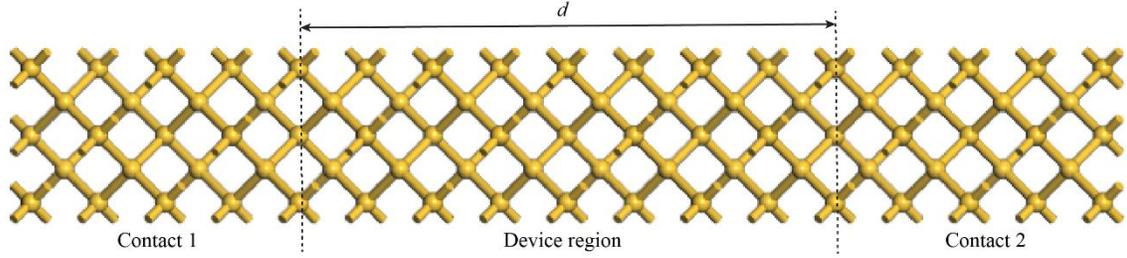

Figure 8. Schematic of the numerical model for phonon NEGF simulation of anharmonic heat transport across a silicon thin film with a thickness $d$. The cross-section period is a conventional unit cell of silicon with 8 atoms, and the total number of atoms in the device region is 8 times the number of unit cells along transport direction.

Table 3. Mesh independence verification for phonon NEGF simulation of anharmonic heat transport across a silicon thin film with a thickness $d$=5uc (1uc=5.4018Å) at 300K.

| Frequency mesh | Transverse wave vector mesh | Ballistic thermal conductance (MW/m$^2$·K) | Thermal conductance (MW/m$^2$·K) |
|---|---|---|---|
| 61 | 4×4 | 1130.18 | 937.73 |
| 81 | 4×4 | 1090.59 | 928.05 |
| 81 | 6×6 | 1069.35 | 894.55 |
| 101 | 4×4 | 1074.60 | 894.32 |
| 101 | 6×6 | 1040.03 | 890.97 |
| 101 | 8×8 | 1053.53 | 872.05 |
| 121 | 6×6 | 1061.24 | 883.33 |
| 121 | 8×8 | 1065.81 | 893.08 |

To provide a benchmark data to the anharmonic phonon NEGF simulation result, we conduct a Monte Carlo (MC) solution of phonon Boltzmann equation for the same cross-plane heat transport with consistent DFT input. In principle, NEGF and Boltzmann formalisms should provide very similar results in the transport regime where the particle picture of phonons is valid and the coherent effects are negligible. For cross-plane heat transport through silicon thin film at room temperature, a recent study has shown that the phonon Boltzmann equation could work down to a thickness of about 10 nm [82]. This critical thickness is sound since the dominant phonon coherence length of silicon at room temperature is about 1 nm [6]. Therefore, for a thickness larger than 10 nm, the Monte Carlo solution could be a good benchmark for the anharmonic phonon NEGF result. We adopt an efficient energy-based deviational phonon MC scheme [83], with the phonon dispersion and relaxation time computed from the same DFT harmonic and anharmonic FCs as that for the phonon NEGF simulation. The phonon MC scheme is solving the Boltzmann equation under the single mode relaxation time (SMRT) scattering term, which is a very good model for silicon with weak normal scattering [9, 80]. As a validation of the present MC code, the thickness-dependent cross-plane thermal conductivity of silicon thin film is shown in Figure 10(a) when the $3^{rd}$ nearest neighboring shell is considered in the third-order DFT FC. The MC result shows a very good agreement with i) previous MC study considering DFT input [84]; ii) a recent experimental measurement [85].

Since only the first nearest neighboring shell is considered in the third-order DFT FC input in the anharmonic phonon NEGF simulation, we also consider the same FC input in MC. The cross-plane thermal conductivity of silicon thin film with a thickness of 20uc (10.8 nm) and 24uc (13 nm) is shown in Table 4. It is seen that the result predicted by the present NEGF simulation is very close to the corresponding MC result. The about 10% underestimation of the NEGF formalism comes from the fact that we still neglect some terms in the diagonal blocks and all the terms in the first off-diagonal blocks of the phonon scattering self-energy as explained in Section 2.3.1. Although these terms are small, including more terms in the scattering self-energy is expected to increase the thermal conductivity a bit, as it will be inferred later in Section 3.3. Taking into account such effect, this comparison can act as a quantitative validation of the present anharmonic phonon NEGF formalism and numerical framework.

Table 4. The cross-plane thermal conductivity of silicon thin film at 300K predicted by Monte Carlo (MC) and by the present anharmonic phonon NEGF with consistent DFT input.

| Film thickness $d$ | Thermal conductivity by MC + DFT (W/m·K) | Thermal conductivity by the present NEGF + DFT (W/m·K) |
|---|---|---|
| 20uc (10.8 nm) | 8.4640 | 7.4489 |
| 24uc (13 nm) | 9.7662 | 8.4025 |

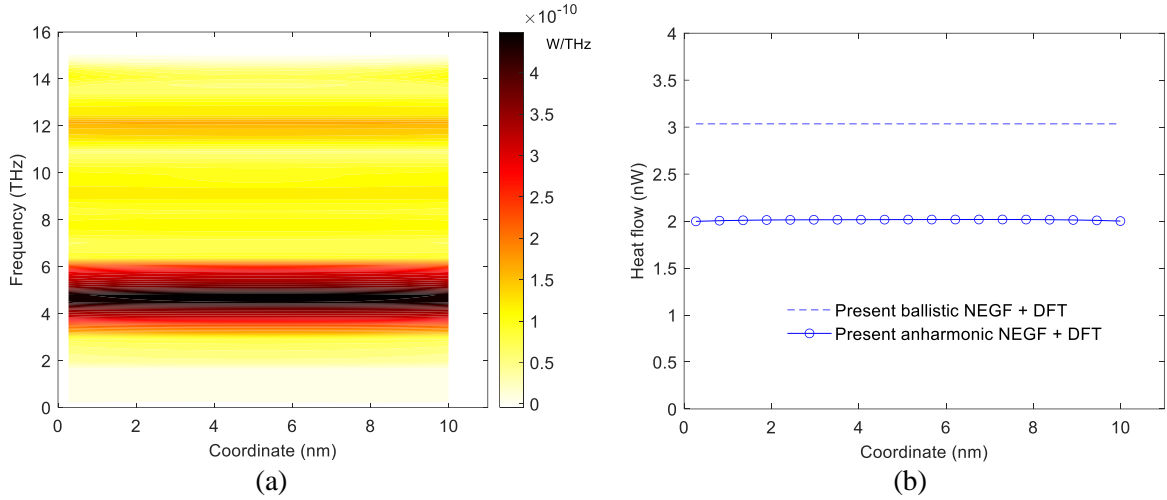

Figure 9. Heat flow across silicon thin film with a thickness of 20uc (10.8nm) at 300K by phonon NEGF: (a) position-dependent spectral heat flow by anharmonic NEGF with DFT input; (b) position-dependent heat flow by ballistic NEGF with DFT input (dashed line) and by anharmonic NEGF with DFT input (solid line with circles).

It is also interesting to note that the results of the phonon Boltzmann equation and the present phonon NEGF formalism agree also well for a film thickness smaller than 10 nm, as shown in Figure 10(b). This indicates that the phonon Boltzmann equation seems to still work for cross-plane heat transport through a silicon thin film of few nanometers. The underlying physical interpretation remains to be investigated in the near future. One possible explanation might be that the coherence of phonons still contributes negligibly to heat transport in this situation. Quantifying the coherence of phonons and its contribution to heat transport remains a open question. Since the computed thermal conductivities have some difference when the first and the third nearest neighboring shell are respectively considered, we show the thickness-dependent cross-plane thermal conductivity normalized by the corresponding bulk value in Figure 11. At small thickness, the present anharmonic phonon NEGF result well agrees with the general trend of non-dimensional thermal conductivity from different reference data.

We also provide the results of local temperature distribution across the silicon thin film with a thickness of 20uc (10.8 nm) and 24uc (13 nm) in Figure 12. The results by the present anharmonic phonon NEGF simulation show a good agreement with those of the MC simulation with the same DFT input. A large temperature jump near the two contacts is obtained due to the strong non-equilibrium effect between the thin film and contacts, which is well known in the heat transport community and has been already reported [6, 7]. This represents a further demonstration of the accuracy of the present formalism and numerical framework. Moreover, the large deviation of local temperature distribution from the result in the ballistic limit indicates that the phonon-phonon scattering is appreciable even at thicknesses around 10 nm. This is also clearly visible in the difference between ballistic and anharmonic results in Figure 10(b), where about 20% reduction of thermal conductivity is obtained due to anharmonic phonon-phonon scattering. Physically this reduction mainly comes from the scattering of optical phonons and LA (longitudinal acoustic) phonons, as is inferred from the spectral heat flow distribution in Figure 13.

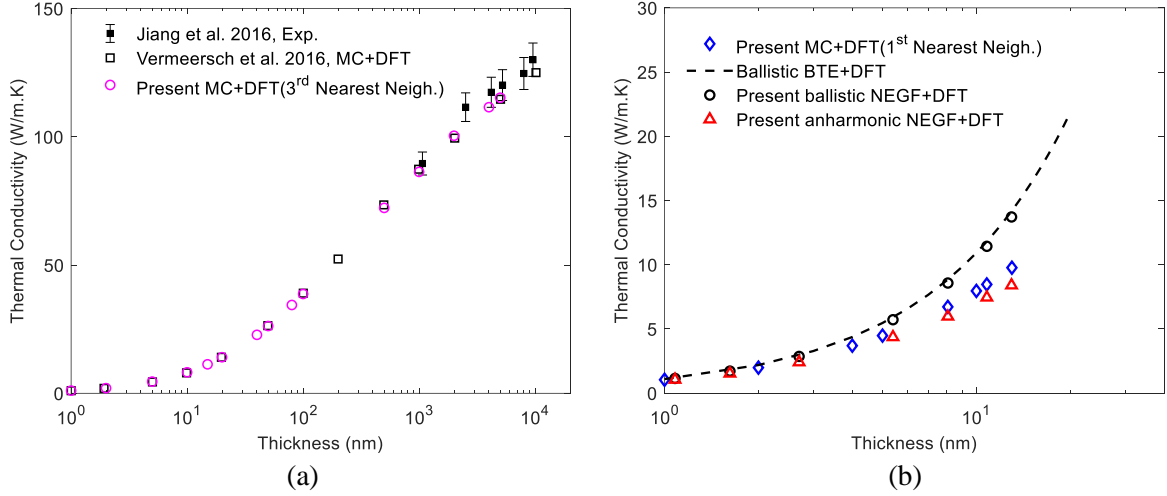

Figure 10. Thickness-dependent cross-plane thermal conductivity of silicon thin film at 300K: (a) comparison of the result by the present Monte Carlo (MC) with DFT input considering 3$^{rd}$ nearest neighboring shell in the third-order FC to previous result by MC with DFT input [84] and experimental data [85]; (b) comparison of the present ballistic and anharmonic NEGF with DFT input with the present MC with DFT input considering 1$^{st}$ nearest neighboring shell in the third-order FC (consistent with that in anharmonic NEGF). The dashed line represents the ballistic solution of Boltzmann transport equation (BTE) with the same DFT input: $\kappa_{\text{eff}} = G_{\text{ballistic}} \cdot d$ where the ballistic thermal conductance is computed by: $G_{\text{ballistic}} = \sum_s \int_{v_{gx}(\mathbf{q},s)>0} v_{gx}(\mathbf{q},s) \hbar\omega(\mathbf{q},s) \partial f^{\text{eq}}/\partial T \, d\mathbf{q}/(2\pi)^3$.

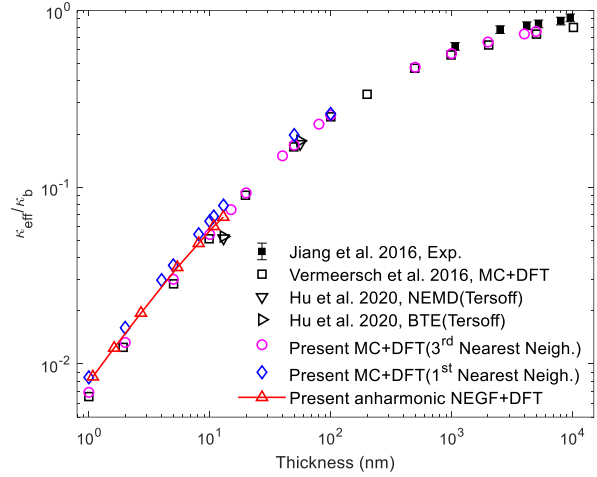

Figure 11. Thickness-dependent non-dimensional cross-plane thermal conductivity of silicon thin film at 300K: experimental data (filled squares with error bar) [85], previous Monte Carlo (MC) with DFT input (squares) [84], non-equilibrium molecular dynamics (NEMD) and Boltzmann transport equation (BTE) with the same Tersoff potential input (triangles) [82], present MC with DFT input considering 3$^{rd}$ (circles) and 1$^{st}$ (diamonds) nearest neighboring shell in the third-order FC, the present anharmonic NEGF with DFT input, considering only 1$^{st}$ nearest neighboring shell in the third-order FC (triangles with line).

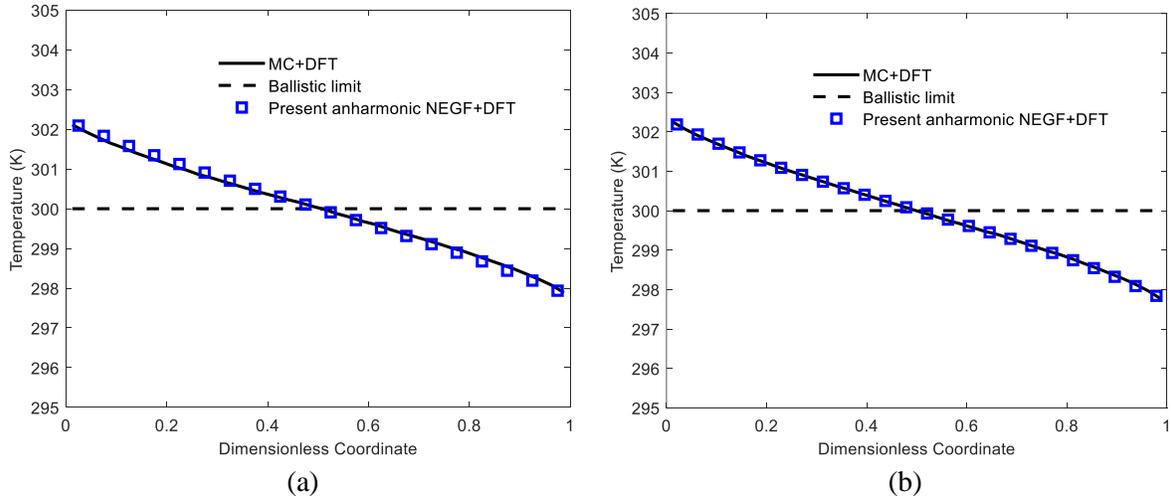

Figure 12. Local temperature distribution in cross-plane heat transport through silicon thin film with different thicknesses under a contact temperature difference of 10K around 300K: (a) $d$=20uc (10.8 nm); (b) $d$=24uc (13 nm). The squares represent the result by the present anharmonic phonon NEGF with DFT input, the solid line represents the result by the present Monte Carlo (MC) with the same DFT input considering only the 1$^{st}$ nearest neighboring shell in the third-order FC, whereas the dashed line represents the uniform distribution in purely ballistic limit.

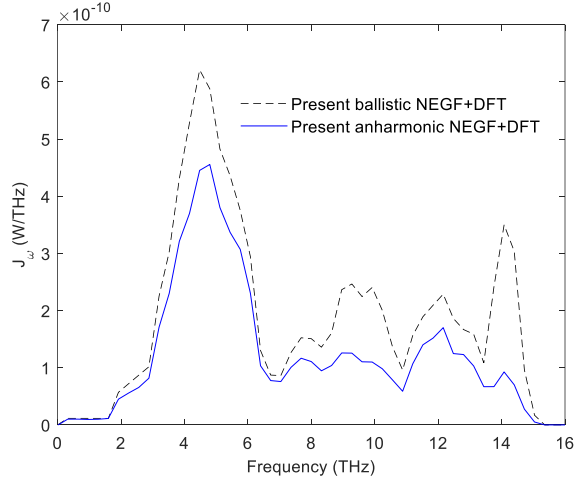

Figure 13. Spectral heat flow in cross-plane heat transport through silicon thin film with a thickness of 20uc (10.8 nm) at 300K: the dashed line denotes the result by ballistic phonon NEGF, the solid line represents the result by anharmonic phonon NEGF. The spectral heat flow has been averaged over the 20 slabs of the device for both cases.

*3.3 Assessment of approximations in numerical implementation*

Finally, we briefly discuss the numerical approximations implemented for the calculation of the anharmonic phonon scattering self-energy, which is crucial to ensure a balance between accuracy and efficiency in large-scale quantum heat transport simulation. As already mentioned in Section 2.3.1, a simpler treatment of the scattering self-energy matrix, by considering only the 3×3 diagonal blocks, is adopted in a previous work by one of the co-authors [64]. Here we also include this approximation into our computational framework and compare the obtained results with those of the present approximation, *i.e.* considering the dominant terms in the $N_x \times N_x$ diagonal blocks (c.f. Section 2.3.1). The heat transport across silicon thin film with two thicknesses of 3uc and 5uc at room temperature is simulated. For the thickness of 3uc, the thermal conductance predicted based on the present approximation and the previous one is respectively: 939.72 and 867.65 (MW/m$^2$·K). For the thickness of 5uc, the thermal conductance based on the present and previous approximation is respectively: 890.97 and 784.09 (MW/m$^2$·K). Comparing to the present approximation, the 3×3 diagonal approximation will underestimate ~8% and ~12% of the thermal conductance of silicon thin film with a thickness of 3uc and 5uc respectively. The underestimation will increase with increasing film thickness due to the stronger phonon-phonon scattering rate. The underlying reason can be understood from the spectral heat flow results shown in Figure 14. It is seen that considering only the 3×3 diagonal blocks of the anharmonic scattering self-energy matrix will significantly overestimate the scattering rate of low-frequency TA (transverse acoustic) phonons which mainly contribute to heat transport. In principle, the incorporation of more terms into the scattering self-energy matrix in the present approximation will help retrieving the underestimation of thermal conductivity shown in Table 4 and Figure 10(b). However, since the dominant terms have been already considered, the present treatment constitutes a good balance between the computational accuracy and efficiency. Note that a further approximation was made in the product of greater/lesser Green's function matrix in the phonon scattering self-energy of Eq. (15) [64]:

$$\Sigma_{s,ll}^{>,<ij}(\omega) = \frac{1}{2}i\hbar \sum_{l_1 l_2} \sum_{j_1 j_2 j_3 j_4} \int_{-\infty}^{\infty} \frac{d\omega'}{2\pi} \Phi_{ll_1 l_2}^{ij_1 j_2} \Phi_{ll_2 l_1}^{jj_3 j_4} G_{l_1 l_1}^{>,<j_1 j_4}(\omega') G_{l_2 l_2}^{>,<j_2 j_3}(\omega - \omega'),\qquad(38)$$

where the matrix product is simplified into the product of their 3×3 diagonal blocks. Such kind of local diagonal approximation is usually assumed in electron NEGF [41, 77] and will speed up the simulation and reduce the computational cost. Yet based on our numerical test, it is a too large simplification for phonon NEGF which will lead to a further overestimation of the phonon-phonon scattering rate and will make the SCBA iteration diverge. The physical interpretation may be that the self-energy for electron-phonon scattering is linearly proportional to the electron Green's function, whereas the self-energy for phonon-phonon scattering is a quadratic function of phonon Green's function, which is a more complicated non-linear problem. This could be a possible explanation for the large scaling of empirical valence-force-field anharmonic FCs to fit the experimental thermal conductivity of silicon therein [64].

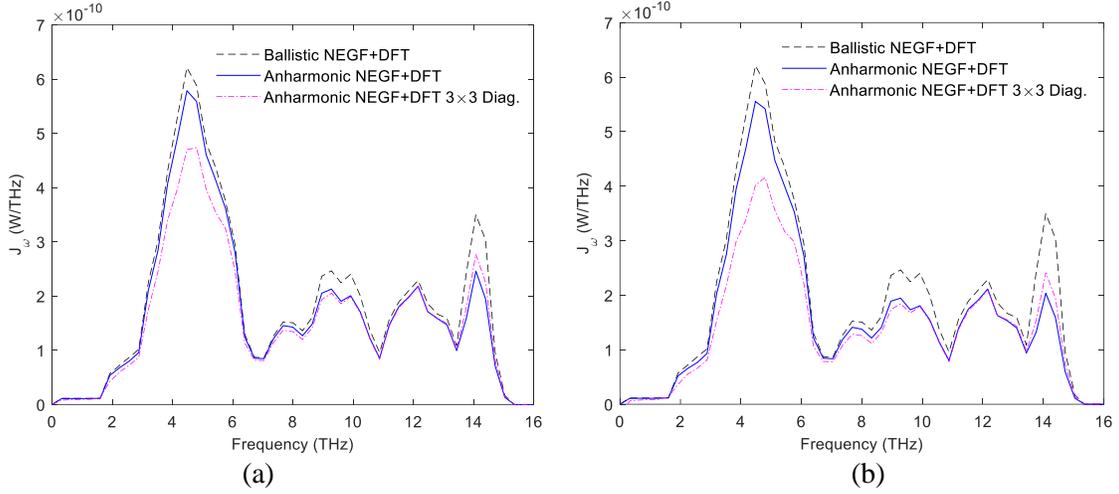

Figure 14. Spectral heat flow in cross-plane heat transport through silicon thin film with different thicknesses at 300K: (a) 3uc, (b) 5uc. The dashed line and solid line represent respectively the result of the present ballistic and anharmonic NEGF with DFT input, the dash-dotted line represents the result of the anharmonic NEGF considering only 3×3 diagonal blocks of the anharmonic phonon scattering self-energy matrix.

## 4. Conclusions

In summary, we present a non-equilibrium Green's function (NEGF) computational framework with the first-principle input for large-scale quantum heat transport simulations accounting for the anharmonic phonon-phonon scattering. The theoretical formulation of anharmonic scattering self-energy is clarified through a careful diagrammatic perturbation analysis, with a Fourier's representation further introduced, which satisfies both the energy and momentum conservation for nanostructures with transverse periodicity. A quantitative validation of the anharmonic phonon NEGF formalism is demonstrated through a comparison to the results by Boltzmann equation in the particle transport regime based on a classical cross-plane heat transport in silicon thin film. The phonon-phonon scattering is shown to be non-negligible even for films with a thickness as small as 10 nm and to introduce a 20% reduction of thermal conductivity at room temperature. The widely used local diagonal approximation of the scattering self-energy in electron NEGF is shown to significantly overestimate the phonon-phonon scattering rates. The present computational framework provides a potential platform for first-principle prediction of thermal boundary conductance at the interface, and for investigation of the transition from coherent to incoherent heat transport in nanophononic crystals like superlattices, which is pending in our future work. This study thus opens new opportunities for the understanding and tuning of phonon heat transport in the coherent quantum wave regime.

**Acknowledgements**

Y. Guo would like to appreciable helpful discussions with S. Fiore from EHT Zürich on the reconstruction of harmonic FC matrix from DFT calculation. This work was supported by the Postdoctoral Fellowship of Japan Society for the Promotion of Science (P19353), and CREST Japan Science and Technology Agency (JPMJCR19I1 and JPMJCR19Q3). This research used the computational resource of the Oakforest-PACS supercomputer system, The University of Tokyo.

**Data Availability**

The data that support the findings of this study are available from the corresponding author upon reasonable request.

**Appendix A. Derivation of anharmonic scattering self-energy through diagrammatic perturbation expansion**

The starting point of the theoretical derivation is the definition of contour-ordered phonon Green's function for heat transport in 1D nanostructures [43, 44]:

$$G_{nm}^{ij}(\tau,\tau') = -i\left\langle T_C\left[u_n^i(\tau)u_m^j(\tau')\right]\right\rangle, \tag{A1}$$

where $T_C$ is the contour ordering operator, and the times $\tau, \tau'$ are on the contour. $u_n^i(\tau)$ and $u_m^j(\tau')$ are atomic displacement operators in the Heisenberg representation. Note in the present derivation we adopt the convention of $\hbar = 1$ [86], which shall be recovered in the final result. Transforming from the Heisenberg representation to the interaction representation of Eq. (A1), we obtain the following expression for the contour-ordered phonon Green's function:

$$G_{nm}^{ij}(\tau,\tau') = -i\left\langle T_C\left[S_C \hat{u}_n^i(\tau)\hat{u}_m^j(\tau')\right]\right\rangle_0, \tag{A2}$$

where $\langle\ \rangle_0$ denotes the expectation over the exactly solvable harmonic part at equilibrium with the density matrix $\rho_0 = \exp(-\beta H_0)/\text{Tr}\left[\exp(-\beta H_0)\right]$, where $\beta = 1/k_B T$. The caret on the operators denotes the interaction representation. The evolution operator in Eq. (A2) is defined as [37, 40]:

$$S_C = \exp\left[-i\int_C d\tau_1 \hat{V}(\tau_1)\right] = \sum_{n=0}^{\infty}\frac{(-i)^n}{n!}\int_C d\tau_1 \int_C d\tau_2 \cdots \int_C d\tau_n \hat{V}(\tau_1)\hat{V}(\tau_2)\cdots\hat{V}(\tau_n), \tag{A3}$$

where the third-order perturbation part of the Hamiltonian in the interaction representation is:

$$\hat{V}(\tau) = \frac{1}{3!}\sum_{nml}\sum_{ijk}\Phi_{nml}^{ijk}\hat{u}_n^i(\tau)\hat{u}_m^j(\tau)\hat{u}_l^k(\tau). \tag{A4}$$

The evolution operator is the basis for the diagrammatic perturbation expansion. Considering the infinite series in Eq. (A3) within second-order and substituting into Eq. (A2), we obtain:

$$G_{nm}^{ij}(\tau,\tau') = G_{nm}^{(0)ij}(\tau,\tau') + G_{nm}^{(1)ij}(\tau,\tau') + G_{nm}^{(2)ij}(\tau,\tau'),\tag{A5}$$

where the zeroth-, first- and second-order terms are respectively:

$$G_{nm}^{(0)ij}(\tau,\tau') = -i\left\langle T_C\left[\hat{u}_n^i(\tau)\hat{u}_m^j(\tau')\right]\right\rangle_0,\tag{A6}$$

$$G_{nm}^{(1)ij}(\tau,\tau') = (-i)^2 \int_C d\tau_1 \left\langle T_C\left[\hat{u}_n^i(\tau)\hat{V}(\tau_1)\hat{u}_m^j(\tau')\right]\right\rangle_0,\tag{A7}$$

$$G_{nm}^{(2)ij}(\tau,\tau') = \frac{(-i)^3}{2}\int_C d\tau_1 \int_C d\tau_2 \left\langle T_C\left[\hat{u}_n^i(\tau)\hat{V}(\tau_1)\hat{V}(\tau_2)\hat{u}_m^j(\tau')\right]\right\rangle_0.\tag{A8}$$

The first-order term in Eq. (A7) will be vanishing since it involves odd number of atomic displacement operators. The full expression of the second-order term will be:

$$\begin{aligned}G_{nm}^{(2)ij}(\tau,\tau') =& \frac{(-i)^3}{2}\int_C d\tau_1\int_C d\tau_2 \frac{1}{3!\cdot 3!}\sum_{n_1 m_1 l_1}\sum_{i_1 j_1 k_1}\sum_{n_2 m_2 l_2}\sum_{i_2 j_2 k_2} \Phi_{n_1 m_1 l_1}^{i_1 j_1 k_1}\Phi_{n_2 m_2 l_2}^{i_2 j_2 k_2}\\ &\times\left\langle T_C\left[\hat{u}_n^i(\tau)\hat{u}_{n_1}^{i_1}(\tau_1)\hat{u}_{m_1}^{j_1}(\tau_1)\hat{u}_{l_1}^{k_1}(\tau_1)\hat{u}_{n_2}^{i_2}(\tau_2)\hat{u}_{m_2}^{j_2}(\tau_2)\hat{u}_{l_2}^{k_2}(\tau_2)\hat{u}_m^j(\tau')\right]\right\rangle_0\end{aligned}.\tag{A9}$$

The Wick's theorem [37, 86] will be applied for the decomposition of the expectation of the product of eight displacement operators, where only the connected diagrams shown in Fig. A1 are considered. These represent the physically feasible three-phonon anharmonic scattering process. There are 3×3×2 pairing combinations for each case in Fig. A1, and totally 36 equivalent pairing combinations such that Eq. (A9) becomes:

$$\begin{aligned}G_{nm}^{(2)ij}(\tau,\tau') =& \frac{(-i)^3}{2}\int_C d\tau_1 \int_C d\tau_2 \frac{36}{3!\cdot 3!}\sum_{n_1 m_1 l_1}\sum_{i_1 j_1 k_1}\sum_{n_2 m_2 l_2}\sum_{i_2 j_2 k_2}\Phi_{n_1 m_1 l_1}^{i_1 j_1 k_1}\Phi_{n_2 m_2 l_2}^{i_2 j_2 k_2}\\ &\times\left\langle T_C\left[\hat{u}_n^i(\tau)\hat{u}_{n_1}^{i_1}(\tau_1)\right]\right\rangle_0\left\langle T_C\left[\hat{u}_{m_1}^{j_1}(\tau_1)\hat{u}_{m_2}^{j_2}(\tau_2)\right]\right\rangle_0\left\langle T_C\left[\hat{u}_{l_1}^{k_1}(\tau_1)\hat{u}_{n_2}^{i_2}(\tau_2)\right]\right\rangle_0\left\langle T_C\left[\hat{u}_{l_2}^{k_2}(\tau_2)\hat{u}_m^j(\tau')\right]\right\rangle_0\end{aligned},\tag{A10}$$

which can be rewritten into:

$$G_{nm}^{(2)ij}(\tau,\tau') = \frac{i}{2}\int_C d\tau_1\int_C d\tau_2 \sum_{n_1 m_1 l_1}\sum_{i_1 j_1 k_1}\sum_{n_2 m_2 l_2}\sum_{i_2 j_2 k_2}\Phi_{n_1 m_1 l_1}^{i_1 j_1 k_1}\Phi_{n_2 m_2 l_2}^{i_2 j_2 k_2} G_{nn_1}^{(0)ii_1}(\tau,\tau_1) G_{m_1 m_2}^{(0)j_1 j_2}(\tau_1,\tau_2) G_{l_1 n_2}^{(0)k_1 i_2}(\tau_1,\tau_2) G_{l_2 m}^{(0)k_2 j}(\tau_2,\tau').$$

(A11)

Putting Eq. (A6) and Eq. (A11) into Eq. (A5), we obtain the contour-order phonon Green's function within second-order as:

$$G_{nm}^{ij}(\tau,\tau') = G_{nm}^{(0)ij}(\tau,\tau') + \frac{i}{2}\int_C d\tau_1 \int_C d\tau_2 \sum_{n_1 m_1 l_1} \sum_{i_1 j_1 k_1} \sum_{n_2 m_2 l_2} \sum_{i_2 j_2 k_2} \Phi_{n_1 m_1 l_1}^{i_1 j_1 k_1} \Phi_{n_2 m_2 l_2}^{i_2 j_2 k_2} \\ \times G_{nn_1}^{(0)ii_1}(\tau,\tau_1) G_{m_1 m_2}^{(0)j_1 j_2}(\tau_1,\tau_2) G_{l_1 n_2}^{(0)k_1 i_2}(\tau_1,\tau_2) G_{l_2 m}^{(0)k_2 j}(\tau_2,\tau') \quad . \tag{A12}$$

Comparing Eq. (A12) to the Dyson's equation it follows [37, 40]:

$$G_{nm}^{ij}(\tau,\tau') = G_{nm}^{(0)ij}(\tau,\tau') + \int_C d\tau_1 \int_C d\tau_2 \sum_{n_1 l_2} \sum_{i_1 k_2} G_{nn_1}^{(0)ii_1}(\tau,\tau_1) \Sigma_{s,n_1 l_2}^{i_1 k_2}(\tau_1,\tau_2) G_{l_2 m}^{k_2 j}(\tau_2,\tau'), \tag{A13}$$

we get the expression of anharmonic phonon scattering self-energy as:

$$\Sigma_{s,nm}^{ij}(\tau_1,\tau_2) = \frac{i}{2} \sum_{m_1 l_1 n_2 m_2} \sum_{j_1 k_1 i_2 j_2} \Phi_{nm_1 l_1}^{ij_1 k_1} \Phi_{mn_2 m_2}^{ji_2 j_2} G_{m_1 m_2}^{j_1 j_2}(\tau_1,\tau_2) G_{l_1 n_2}^{k_1 i_2}(\tau_1,\tau_2), \tag{A14}$$

where the unperturbed Green's function has been replaced by the full Green's function based on the self-consistent Born approximation.

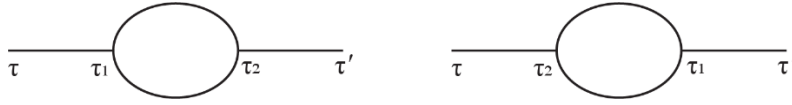

Fig. A1. Connected Feynman diagrams for three-phonon anharmonic scattering process

Through the analytic continuation process based on the Langreth theorem [40], we get the greater/lesser scattering self-energy in real time from Eq. (A14):

$$\Sigma_{s,nm}^{>,<ij}(t_1,t_2) = \frac{i}{2} \sum_{m_1 l_1 n_2 m_2} \sum_{j_1 k_1 i_2 j_2} \Phi_{nm_1 l_1}^{ij_1 k_1} \Phi_{mn_2 m_2}^{ji_2 j_2} G_{m_1 m_2}^{>,<j_1 j_2}(t_1,t_2) G_{l_1 n_2}^{>,<k_1 i_2}(t_1,t_2). \tag{A15}$$

For the stationary state heat transport considered in this work, the dependence on $t_1$ and $t_2$ will be reduced to the dependence on $(t_1-t_2)$. For convenience, the following Fourier transform and its inverse transform is introduced [86]:

$$\Sigma_{s,nm}^{>,<ij}(t_1,t_2) = \Sigma_{s,nm}^{>,<ij}(t_1-t_2) = \int_{-\infty}^{\infty} \frac{d\omega}{2\pi} \exp\left[-i(t_1-t_2)\omega\right] \Sigma_{s,nm}^{>,<ij}(\omega), \tag{A16}$$

$$\Sigma_{s,nm}^{>,<ij}(\omega) = \int_{-\infty}^{\infty} dt_1 \exp\left[i\omega(t_1-t_2)\right] \Sigma_{s,nm}^{>,<ij}(t_1-t_2). \tag{A17}$$

Fourier transforms of Eqs. (A16) and (A17) are also applicable to the Green's function in Eq. (A15). After the Fourier transform, Eq. (A15) becomes:

$$\Sigma_{s,nm}^{>,<ij}(\omega) = \frac{i}{2} \sum_{m_1 l_1 n_2 m_2} \sum_{j_1 k_1 i_2 j_2} \int_{-\infty}^{\infty} \frac{d\omega'}{2\pi} \Phi_{nm_1l_1}^{ij_1k_1} \Phi_{mn_2m_2}^{ji_2j_2} G_{m_1m_2}^{>,<j_1j_2}(\omega') G_{l_1n_2}^{>,<k_1i_2}(\omega-\omega'), \tag{A18}$$

which is exactly Eq. (15) in the main text. It is also seen from Eq. (A18) that the energy conservation is automatically satisfied in the three-phonon scattering process: $\omega = \omega' + (\omega - \omega')$.

**Appendix B. Fourier's representation of the anharmonic phonon-phonon scattering self-energy matrix**

The starting point of the derivation is the expression of anharmonic phonon scattering self-energy matrix Eq. (15) for 1D nanostructures:

$$\Sigma_{s,ll'}^{>,<ij}(\omega) = \frac{1}{2} i\hbar \sum_{l_1l_2l_3l_4} \sum_{j_1j_2j_3j_4} \int_{-\infty}^{\infty} \frac{d\omega'}{2\pi} \Phi_{ll_1l_2}^{ij_1j_2} \Phi_{l'l_3l_4}^{jj_3j_4} G_{l_1l_4}^{>,<j_1j_4}(\omega') G_{l_2l_3}^{>,<j_2j_3}(\omega-\omega'). \tag{B1}$$

The inverse transform of the Fourier's representation Eq. (23) for the anharmonic FC matrix is:

$$\Phi_{ll_1l_2}^{ij_1j_2} = \frac{1}{N^2} \sum_{\mathbf{q}_\perp} \sum_{\mathbf{q}'_\perp} \tilde{\Phi}_{l_xl_{1x}l_{2x}}^{ij_1j_2}(\mathbf{q}_\perp,\mathbf{q}'_\perp) \exp(i\mathbf{q}_\perp \cdot \Delta\mathbf{R}_\perp) \exp(i\mathbf{q}'_\perp \cdot \Delta\mathbf{R}'_\perp). \tag{B2}$$

For convenience of later derivation, Eq. (B2) is slightly rewritten into:

$$\Phi_{ll_1l_2}^{ij_1j_2} = \frac{1}{N^2} \sum_{\mathbf{q}_\perp} \sum_{\mathbf{q}'_\perp} \tilde{\Phi}_{l_xl_{1x}l_{2x}}^{ij_1j_2}(\mathbf{q}_\perp,\mathbf{q}'_\perp) \exp\left[i(\mathbf{q}_\perp+\mathbf{q}'_\perp) \cdot \Delta\mathbf{R}_\perp\right] \exp(i\mathbf{q}'_\perp \cdot \Delta\mathbf{R}''_\perp), \tag{B3}$$

with $\Delta\mathbf{R}''_\perp = \Delta\mathbf{R}'_\perp - \Delta\mathbf{R}_\perp = (l_{1y}-l_{2y})\mathbf{a}_2 + (l_{1z}-l_{2z})\mathbf{a}_3$.

The Fourier's representation of the scattering self-energy, Green's function and third-order FC matrix in Eq. (B1) can be expressed as below:

$$\Sigma_{s,ll'}^{<,>ij}(\omega) = \frac{1}{N}\sum_{\mathbf{q}}\exp(i\mathbf{q}\cdot\Delta\mathbf{R})\Sigma_{s,l_x l'_x}^{<,>ij}(\omega;\mathbf{q}), \tag{B4}$$

$$G_{l_1 l_4}^{<,>j_1 j_4}(\omega') = \frac{1}{N}\sum_{\mathbf{q}_1}\exp(i\mathbf{q}_1\cdot\Delta\mathbf{R}_1)G_{l_{1x}l_{4x}}^{<,>j_1 j_4}(\omega';\mathbf{q}_1), \tag{B5}$$

$$G_{l_2 l_3}^{<,>j_2 j_3}(\omega-\omega') = \frac{1}{N}\sum_{\mathbf{q}_2}\exp(i\mathbf{q}_2\cdot\Delta\mathbf{R}_2)G_{l_{2x}l_{3x}}^{<,>j_2 j_3}(\omega-\omega';\mathbf{q}_2), \tag{B6}$$

$$\Phi_{l l_1 l_2}^{ij_1 j_2} = \frac{1}{N^2}\sum_{\mathbf{q}_3}\sum_{\mathbf{q}_4}\tilde{\Phi}_{l_x l_{1x} l_{2x}}^{ij_1 j_2}(\mathbf{q}_3,\mathbf{q}_4)\exp\left[i(\mathbf{q}_3+\mathbf{q}_4)\cdot\Delta\mathbf{R}_3\right]\exp(i\mathbf{q}_4\cdot\Delta\mathbf{R}_4), \tag{B7}$$

$$\Phi_{l' l_3 l_4}^{j j_3 j_4} = \frac{1}{N^2}\sum_{\mathbf{q}_5}\sum_{\mathbf{q}_6}\tilde{\Phi}_{l'_x l_{3x} l_{4x}}^{j j_3 j_4}(-\mathbf{q}_5,-\mathbf{q}_6)\exp\left[i(\mathbf{q}_5+\mathbf{q}_6)\cdot\Delta\mathbf{R}_5\right]\exp(i\mathbf{q}_6\cdot\Delta\mathbf{R}_6), \tag{B8}$$

where all the subscripts '$\perp$' are omitted to keep a simple notation, and the relative displacements are defined as below:

$$\Delta\mathbf{R} = (l_y - l'_y)\mathbf{a}_2 + (l_z - l'_z)\mathbf{a}_3, \tag{B9}$$

$$\Delta\mathbf{R}_1 = (l_{1y} - l_{4y})\mathbf{a}_2 + (l_{1z} - l_{4z})\mathbf{a}_3, \tag{B10}$$

$$\Delta\mathbf{R}_2 = (l_{2y} - l_{3y})\mathbf{a}_2 + (l_{2z} - l_{3z})\mathbf{a}_3, \tag{B11}$$

$$\Delta\mathbf{R}_3 = (l_y - l_{1y})\mathbf{a}_2 + (l_z - l_{1z})\mathbf{a}_3, \tag{B12}$$

$$\Delta\mathbf{R}_4 = (l_{1y} - l_{2y})\mathbf{a}_2 + (l_{1z} - l_{2z})\mathbf{a}_3, \tag{B13}$$

$$\Delta\mathbf{R}_5 = (l_{3y} - l'_y)\mathbf{a}_2 + (l_{3z} - l'_z)\mathbf{a}_3, \tag{B14}$$

$$\Delta\mathbf{R}_6 = (l_{4y} - l_{3y})\mathbf{a}_2 + (l_{4z} - l_{3z})\mathbf{a}_3. \tag{B15}$$

Substitution of Eqs. (B4)-(B8) into Eq. (B1) gives rise to:

$$\frac{1}{N}\sum_{\mathbf{q}}\exp(i\mathbf{q}\cdot\Delta\mathbf{R})\Sigma_{s,l_xl'_x}^{<,>ij}(\omega;\mathbf{q}) = \frac{i\hbar}{2}\sum_{l_1l_2l_3l_4}\sum_{j_1j_2j_3j_4}\frac{1}{N^2}\sum_{\mathbf{q}_3}\sum_{\mathbf{q}_4}\tilde{\Phi}_{l_xl_{1x}l_{2x}}^{ij_1j_2}(\mathbf{q}_3,\mathbf{q}_4)\exp\left[i(\mathbf{q}_3+\mathbf{q}_4)\cdot\Delta\mathbf{R}_3\right]\exp(i\mathbf{q}_4\cdot\Delta\mathbf{R}_4)$$

$$\times\frac{1}{N^2}\sum_{\mathbf{q}_5}\sum_{\mathbf{q}_6}\tilde{\Phi}_{l'_xl_{3x}l_{4x}}^{jj_3j_4}(-\mathbf{q}_5,-\mathbf{q}_6)\exp\left[i(\mathbf{q}_5+\mathbf{q}_6)\cdot\Delta\mathbf{R}_5\right]\exp(i\mathbf{q}_6\cdot\Delta\mathbf{R}_6)$$

$$\times\int_{-\infty}^{\infty}\frac{d\omega'}{2\pi}\frac{1}{N}\sum_{\mathbf{q}_1}\exp(i\mathbf{q}_1\cdot\Delta\mathbf{R}_1)G_{l_{1x}l_{4x}}^{<,>j_1j_4}(\omega';\mathbf{q}_1)\frac{1}{N}\sum_{\mathbf{q}_2}\exp(i\mathbf{q}_2\cdot\Delta\mathbf{R}_2)G_{l_{2x}l_{3x}}^{<,>j_2j_3}(\omega-\omega';\mathbf{q}_2)$$

(B16)

Since $\Delta\mathbf{R} = \Delta\mathbf{R}_3 + \Delta\mathbf{R}_4 + \Delta\mathbf{R}_2 + \Delta\mathbf{R}_5$, multiplying $\exp(-i\mathbf{q}'\cdot\Delta\mathbf{R})$ on both sides of Eq. (B16) and summing over $(l_y, l_z)$, we reform the left-hand side term as:

$$\frac{1}{N}\sum_{\mathbf{q}}\sum_{\Delta\mathbf{R}}\exp\left[i(\mathbf{q}-\mathbf{q}')\cdot\Delta\mathbf{R}\right]\Sigma_{s,l_xl'_x}^{<,>ij}(\omega;\mathbf{q}) = \Sigma_{s,l_xl'_x}^{<,>ij}(\omega;\mathbf{q}'),$$

(B17)

where a classical relation in lattice dynamics has been used [8, 73]:

$$\sum_{\Delta\mathbf{R}}\exp\left[i(\mathbf{q}-\mathbf{q}')\cdot\Delta\mathbf{R}\right] = N\Delta(\mathbf{q}-\mathbf{q}').$$

(B18)

The right-hand side term becomes:

$$\frac{i\hbar}{2}\sum_{l_1l_2l_3l_4}\sum_{j_1j_2j_3j_4}\frac{1}{N^2}\sum_{\mathbf{q}_3}\sum_{\mathbf{q}_4}\tilde{\Phi}_{l_xl_{1x}l_{2x}}^{ij_1j_2}(\mathbf{q}_3,\mathbf{q}_4)\sum_{\Delta\mathbf{R}_3}\exp\left[i(\mathbf{q}_3+\mathbf{q}_4-\mathbf{q}')\cdot\Delta\mathbf{R}_3\right]\exp\left[i(\mathbf{q}_4-\mathbf{q}')\cdot\Delta\mathbf{R}_4\right]$$

$$\times\frac{1}{N^2}\sum_{\mathbf{q}_5}\sum_{\mathbf{q}_6}\tilde{\Phi}_{l'_xl_{3x}l_{4x}}^{jj_3j_4}(-\mathbf{q}_5,-\mathbf{q}_6)\exp\left[i(\mathbf{q}_5+\mathbf{q}_6-\mathbf{q}')\cdot\Delta\mathbf{R}_5\right]\exp(i\mathbf{q}_6\cdot\Delta\mathbf{R}_6)$$

$$\times\int_{-\infty}^{\infty}\frac{d\omega'}{2\pi}\frac{1}{N}\sum_{\mathbf{q}_1}\exp(i\mathbf{q}_1\cdot\Delta\mathbf{R}_1)G_{l_{1x}l_{4x}}^{<,>j_1j_4}(\omega';\mathbf{q}_1)\frac{1}{N}\sum_{\mathbf{q}_2}\exp\left[i(\mathbf{q}_2-\mathbf{q}')\cdot\Delta\mathbf{R}_2\right]G_{l_{2x}l_{3x}}^{<,>j_2j_3}(\omega-\omega';\mathbf{q}_2)$$

$$=\frac{i\hbar}{2}\sum_{l_1l_2l_3l_4}\sum_{j_1j_2j_3j_4}\frac{1}{N}\sum_{\mathbf{q}_4}\tilde{\Phi}_{l_xl_{1x}l_{2x}}^{ij_1j_2}(\mathbf{q}'-\mathbf{q}_4,\mathbf{q}_4)\exp\left[i(\mathbf{q}_4-\mathbf{q}')\cdot\Delta\mathbf{R}_4\right]$$

$$\times\frac{1}{N^2}\sum_{\mathbf{q}_5}\sum_{\mathbf{q}_6}\tilde{\Phi}_{l'_xl_{3x}l_{4x}}^{jj_3j_4}(-\mathbf{q}_5,-\mathbf{q}_6)\exp\left[i(\mathbf{q}_5+\mathbf{q}_6-\mathbf{q}')\cdot\Delta\mathbf{R}_5\right]\exp(i\mathbf{q}_6\cdot\Delta\mathbf{R}_6)$$

$$\times\int_{-\infty}^{\infty}\frac{d\omega'}{2\pi}\frac{1}{N}\sum_{\mathbf{q}_1}\exp(i\mathbf{q}_1\cdot\Delta\mathbf{R}_1)G_{l_{1x}l_{4x}}^{<,>j_1j_4}(\omega';\mathbf{q}_1)\frac{1}{N}\sum_{\mathbf{q}_2}\exp\left[i(\mathbf{q}_2-\mathbf{q}')\cdot\Delta\mathbf{R}_2\right]G_{l_{2x}l_{3x}}^{<,>j_2j_3}(\omega-\omega';\mathbf{q}_2)$$

(B19)

In the derivation of Eq. (B19), we have identified the sum over $(l_y, l_z)$ as a sum over $\Delta \mathbf{R}_3$, and also used the relation Eq. (B18). Therefore, the scattering self-energy is:

$$\Sigma_{s, l_x l'_x}^{<,>ij}(\omega; \mathbf{q}') = \frac{i\hbar}{2} \sum_{l_1 l_2 l_3 l_4} \sum_{j_1 j_2 j_3 j_4} \frac{1}{N} \sum_{\mathbf{q}_4} \tilde{\Phi}_{l_x l_{1x} l_{2x}}^{ij_1 j_2}(\mathbf{q}' - \mathbf{q}_4, \mathbf{q}_4) \exp[i(\mathbf{q}_4 - \mathbf{q}') \cdot \Delta \mathbf{R}_4]$$
$$\times \frac{1}{N^2} \sum_{\mathbf{q}_5} \sum_{\mathbf{q}_6} \tilde{\Phi}_{l'_x l_{3x} l_{4x}}^{jj_3 j_4}(-\mathbf{q}_5, -\mathbf{q}_6) \exp[i(\mathbf{q}_5 + \mathbf{q}_6 - \mathbf{q}') \cdot \Delta \mathbf{R}_5] \exp(i\mathbf{q}_6 \cdot \Delta \mathbf{R}_6)$$
$$\times \int_{-\infty}^{\infty} \frac{d\omega'}{2\pi} \frac{1}{N} \sum_{\mathbf{q}_1} \exp(i\mathbf{q}_1 \cdot \Delta \mathbf{R}_1) G_{l_{1x} l_{4x}}^{<,>j_1 j_4}(\omega'; \mathbf{q}_1) \frac{1}{N} \sum_{\mathbf{q}_2} \exp[i(\mathbf{q}_2 - \mathbf{q}') \cdot \Delta \mathbf{R}_2] G_{l_{2x} l_{3x}}^{<,>j_2 j_3}(\omega - \omega'; \mathbf{q}_2)$$

(B20)

Since only $\Delta \mathbf{R}_5$ depends on $(l'_y, l'_z)$, summing both sides of Eq. (B20) over $(l'_y, l'_z)$ (identified as a sum over $\Delta \mathbf{R}_5$) and using again the relation Eq. (B18), we could reduce Eq. (B20) to:

$$N \Sigma_{s, l_x l'_x}^{<,>ij}(\omega; \mathbf{q}')$$
$$= \frac{i\hbar}{2} \sum_{l_1 l_2 l_3 l_4} \sum_{j_1 j_2 j_3 j_4} \frac{1}{N} \sum_{\mathbf{q}_4} \tilde{\Phi}_{l_x l_{1x} l_{2x}}^{ij_1 j_2}(\mathbf{q}' - \mathbf{q}_4, \mathbf{q}_4) \exp[i(\mathbf{q}_4 - \mathbf{q}') \cdot \Delta \mathbf{R}_4] \frac{1}{N} \sum_{\mathbf{q}_6} \tilde{\Phi}_{l'_x l_{3x} l_{4x}}^{jj_3 j_4}(\mathbf{q}_6 - \mathbf{q}', -\mathbf{q}_6) \exp(i\mathbf{q}_6 \cdot \Delta \mathbf{R}_6).$$
$$\times \int_{-\infty}^{\infty} \frac{d\omega'}{2\pi} \frac{1}{N} \sum_{\mathbf{q}_1} \exp(i\mathbf{q}_1 \cdot \Delta \mathbf{R}_1) G_{l_{1x} l_{4x}}^{<,>j_1 j_4}(\omega'; \mathbf{q}_1) \frac{1}{N} \sum_{\mathbf{q}_2} \exp[i(\mathbf{q}_2 - \mathbf{q}') \cdot \Delta \mathbf{R}_2] G_{l_{2x} l_{3x}}^{<,>j_2 j_3}(\omega - \omega'; \mathbf{q}_2)$$

(B21)

Since $\Delta \mathbf{R}_4 = \Delta \mathbf{R}_1 - \Delta \mathbf{R}_2 + \Delta \mathbf{R}_6$, Eq. (B21) is reformulated into:

$$N \Sigma_{s, l_x l'_x}^{<,>ij}(\omega; \mathbf{q}') = \frac{i\hbar}{2} \sum_{l_1 l_2 l_3 l_4} \sum_{j_1 j_2 j_3 j_4} \frac{1}{N} \sum_{\mathbf{q}_4} \tilde{\Phi}_{l_x l_{1x} l_{2x}}^{ij_1 j_2}(\mathbf{q}' - \mathbf{q}_4, \mathbf{q}_4) \frac{1}{N} \sum_{\mathbf{q}_6} \tilde{\Phi}_{l'_x l_{3x} l_{4x}}^{jj_3 j_4}(\mathbf{q}_6 - \mathbf{q}', -\mathbf{q}_6) \exp[i(\mathbf{q}_6 + \mathbf{q}_4 - \mathbf{q}') \cdot \Delta \mathbf{R}_6]$$
$$\times \int_{-\infty}^{\infty} \frac{d\omega'}{2\pi} \frac{1}{N} \sum_{\mathbf{q}_1} \exp[i(\mathbf{q}_1 + \mathbf{q}_4 - \mathbf{q}') \cdot \Delta \mathbf{R}_1] G_{l_{1x} l_{4x}}^{<,>j_1 j_4}(\omega'; \mathbf{q}_1) \frac{1}{N} \sum_{\mathbf{q}_2} \exp[i(\mathbf{q}_2 - \mathbf{q}_4) \cdot \Delta \mathbf{R}_2] G_{l_{2x} l_{3x}}^{<,>j_2 j_3}(\omega - \omega'; \mathbf{q}_2)$$

(B22)

In this way, the sum over $(l_{1y}, l_{1z})$ (identified as a sum over $\Delta \mathbf{R}_1$) and the sum over $(l_{2y}, l_{2z})$ (identified as a sum over $\Delta \mathbf{R}_2$) become independent, and the relation Eq. (B18) is used to remove the dependence on $\Delta \mathbf{R}_1$ and $\Delta \mathbf{R}_2$. Then the sum over $(l_{3y}, l_{3z})$ (identified as a sum over $\Delta \mathbf{R}_6$) becomes independent and the relation Eq. (B18) is used again. Finally, Eq. (B22) will be reduced to the following expression:

$$\Sigma_{s,l_xl_x'}^{<,>ij}(\omega;\mathbf{q}') = \frac{i\hbar}{2} \sum_{l_{1x}l_{2x}l_{3x}l_{4x}} \sum_{j_1j_2j_3j_4} \frac{1}{N}\sum_{\mathbf{q}_4} \tilde{\Phi}_{l_xl_{1x}l_{2x}}^{ij_1j_2}(\mathbf{q}'-\mathbf{q}_4,\mathbf{q}_4) \tilde{\Phi}_{l_x'l_{3x}l_{4x}}^{jj_3j_4}(-\mathbf{q}_4,\mathbf{q}_4-\mathbf{q}') \\ \times \int_{-\infty}^{\infty} \frac{d\omega'}{2\pi} G_{l_{1x}l_{4x}}^{<,>j_1j_4}(\omega';\mathbf{q}'-\mathbf{q}_4) G_{l_{2x}l_{3x}}^{<,>j_2j_3}(\omega-\omega';\mathbf{q}_4)$$ (B23)

Introducing a variable change: $\mathbf{q}'' = \mathbf{q}' - \mathbf{q}_4$ and rearranging the notation ($\mathbf{q}' \rightarrow \mathbf{q}, \mathbf{q}'' \rightarrow \mathbf{q}'$), we get the final expression of anharmonic phonon-phonon scattering self-energy in the Fourier's representation as:

$$\Sigma_{s,l_xl_x'}^{<,>ij}(\omega;\mathbf{q}) = \frac{1}{2}i\hbar \sum_{l_{1x}l_{2x}l_{3x}l_{4x}} \sum_{j_1j_2j_3j_4} \frac{1}{N}\sum_{\mathbf{q}'} \tilde{\Phi}_{l_xl_{1x}l_{2x}}^{ij_1j_2}(\mathbf{q}',\mathbf{q}-\mathbf{q}') \tilde{\Phi}_{l_x'l_{3x}l_{4x}}^{jj_3j_4}(\mathbf{q}'-\mathbf{q},-\mathbf{q}') \\ \times \int_{-\infty}^{\infty} \frac{d\omega'}{2\pi} G_{l_{1x}l_{4x}}^{<,>j_1j_4}(\omega';\mathbf{q}') G_{l_{2x}l_{3x}}^{<,>j_2j_3}(\omega-\omega';\mathbf{q}-\mathbf{q}')$$ (B24)

**Appendix C. Proof of a general relation between non-equilibrium phonon Green's functions for 3D nanostructures**

Based on the definition of Fourier's representation in Eq. (17), the greater/lesser phonon Green's functions are expressed respectively as:

$$G_{l_xl_x'}^{>,ij}(\omega;\mathbf{q}_\perp) = \sum_{\Delta\mathbf{R}_\perp} \exp(-i\mathbf{q}_\perp \cdot \Delta\mathbf{R}_\perp) G_{ll'}^{>,ij}(\omega),$$ (C1)

$$G_{l_xl_x'}^{<,ij}(\omega;\mathbf{q}_\perp) = \sum_{\Delta\mathbf{R}_\perp} \exp(-i\mathbf{q}_\perp \cdot \Delta\mathbf{R}_\perp) G_{ll'}^{<,ij}(\omega).$$ (C2)

The element of the transpose of the greater Green's function matrix is:

$$\left[\mathbf{G}^>(\omega;\mathbf{q}_\perp)\right]_{l_xl_x'}^{T,ij} = G_{l_x'l_x}^{>,ji}(\omega;\mathbf{q}_\perp) = \sum_{\Delta\mathbf{R}_\perp'} \exp(-i\mathbf{q}_\perp \cdot \Delta\mathbf{R}_\perp') G_{l'l}^{>,ji}(\omega),$$ (C3)

with $\Delta\mathbf{R}_\perp' = (l_y' - l_y)\mathbf{a}_2 + (l_z' - l_z)\mathbf{a}_3$ here. The following symmetry relation is valid between the greater/lesser phonon Green's function for 1D nanostructures without any periodicity [62]:

$$G_{l'l}^{>,ji}(\omega) = G_{ll'}^{<,ij}(-\omega).$$ (C4)

Substituting Eq. (C4) into Eq. (C3), we obtain:

$$\left[\mathbf{G}^{>}(\omega;\mathbf{q}_{\perp})\right]_{l_x l_x'}^{\mathrm{T},ij} = \sum_{\Delta \mathbf{R}_{\perp}'} \exp(-i\mathbf{q}_{\perp}\cdot\Delta\mathbf{R}_{\perp}') G_{ll'}^{<,ij}(-\omega). \tag{C5}$$

Since we have $\Delta\mathbf{R}_{\perp}' = -\Delta\mathbf{R}_{\perp}$, Eq. (C5) can be rewritten into:

$$\left[\mathbf{G}^{>}(\omega;\mathbf{q}_{\perp})\right]_{l_x l_x'}^{\mathrm{T},ij} = \sum_{\Delta \mathbf{R}_{\perp}} \exp(i\mathbf{q}_{\perp}\cdot\Delta\mathbf{R}_{\perp}) G_{ll'}^{<,ij}(-\omega). \tag{C6}$$

Based on the definition in Eq. (C2), Eq. (C6) becomes exactly:

$$\left[\mathbf{G}^{>}(\omega;\mathbf{q}_{\perp})\right]_{l_x l_x'}^{\mathrm{T},ij} = G_{l_x l_x'}^{<,ij}(-\omega;-\mathbf{q}_{\perp}), \tag{C7}$$

which can be reformulated into matrix form as:

$$\left[\mathbf{G}^{>}(\omega;\mathbf{q}_{\perp})\right]^{\mathrm{T}} = \mathbf{G}^{<}(-\omega;-\mathbf{q}_{\perp}). \tag{C8}$$

**Appendix D. MPI scheme for parallelized calculation of anharmonic phonon scattering self-energy**

We follow the basic procedure in the MPI scheme for the parallelized calculation of electron-phonon scattering self-energy matrix in electron NEGF [76]. However, the present situation is slightly more complicated since for each mode $(\omega,\mathbf{q}_{\perp})$ we have to consider the Green's functions of two other modes $(\omega';\mathbf{q}_{\perp}')$ and $(\omega-\omega';\mathbf{q}_{\perp}-\mathbf{q}_{\perp}')$ due to the three-phonon anharmonic scattering process shown in Figure 2. As a first step, for each mode $(\omega,\mathbf{q}_{\perp})$ in a CPU, we build a connection table which stores the information of all the possible connected modes $(\omega-\omega';\mathbf{q}_{\perp}-\mathbf{q}_{\perp}')$ via all possible modes $(\omega';\mathbf{q}_{\perp}')$. The data exchange is then conducted based on the connection table. Since each mode $(\omega,\mathbf{q}_{\perp})$ requires the data of mode $(\omega';\mathbf{q}_{\perp}')$ and mode $(\omega-\omega';\mathbf{q}_{\perp}-\mathbf{q}_{\perp}')$, the data exchange consists of a sending sub-step and a receiving sub-step successively: (I) for all the CPUs, send the local data of mode $(\omega,\mathbf{q}_{\perp})$ to the CPUs corresponding to all the possible modes $(\omega';\mathbf{q}_{\perp}')$, and also to the CPUs corresponding to all the possible modes $(\omega+\omega';\mathbf{q}_{\perp}+\mathbf{q}_{\perp}')$; (II) for all the CPUs, receive the data of all the possible modes $(\omega'; \mathbf{q}'_\perp)$ from the corresponding CPUs, and also the data of all the possible modes $(\omega-\omega'; \mathbf{q}_\perp - \mathbf{q}'_\perp)$ from the corresponding CPUs.

Another important issue is the storage of third-order FC matrix in the Fourier's representation $\tilde{\Phi}^{ij_1j_2}_{l_x l_{1x} l_{2x}}(\mathbf{q}_\perp, \mathbf{q}'_\perp)$, which will be an extremely large matrix for long nanostructures due to its dependence on two transverse wave vectors. We reduce the memory cost by storing its dependence on only the first wave vector in a local CPU attributed to the full parallelization of transverse wave vector (the local wave vector acting as the second wave vector). In this way, we also need to exchange data of the third-order FC matrix when computing the anharmonic scattering self-energy Eq. (21). The idea and procedure are very similar to that for the data exchange of Green's function and are not repeated here for elegance. Once the data exchange of both Green's function and third-order FC matrix is accomplished, the scattering self-energy for each mode $(\omega, \mathbf{q}_\perp)$ can be then calculated based on Eq. (21).

The scalability of the present MPI parallelization scheme is demonsrated in Fig. D1, which shows a scaling of computational time cost versus number of CPUs very close to the ideal scaling limit.

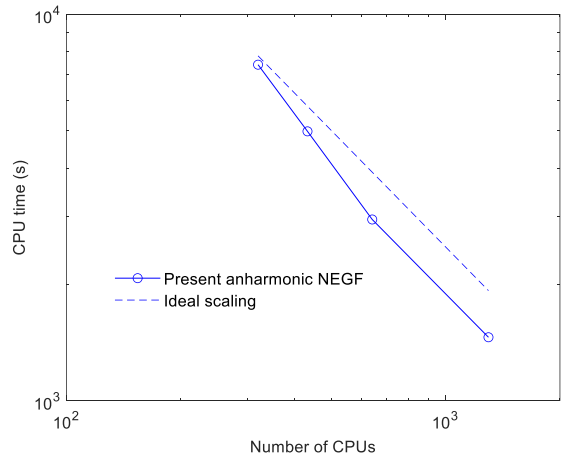

Fig. D1. Computational time cost versus the number of CPUs: the solid line with circles represent the present anharmonic phonon NEGF simulation of heat transport across silicon thin film with a thickness of 2uc at 300K, a frequency mesh of $N_m$=81 and tranverse wave vector mesh of 4×4 is adopted for test. The dashed line represents the ideal scaling.